\documentclass[12pt]{article}
\usepackage[a4paper]{geometry}
\usepackage[italian,english]{babel}

\usepackage{amsmath}
\usepackage{amsfonts}
\usepackage{amstext}
\usepackage{amssymb}
\usepackage{amsthm}
\usepackage{amscd}

\usepackage{hyperref}
\hypersetup{colorlinks,
linkcolor=myrefcolor,
citecolor=mycitecolor,
urlcolor=myurlcolor}
\usepackage[capitalize]{cleveref}
\usepackage{caption}
\usepackage{etaremune}
\usepackage{bibentry}

\usepackage{xcolor}
\definecolor{myurlcolor}{rgb}{0,0,0.4}
\definecolor{mycitecolor}{rgb}{0,0.5,0}
\definecolor{myrefcolor}{rgb}{0.5,0,0}
\usepackage{graphicx}
\usepackage{tikz}
\usepackage{tikz-cd}
\usepackage{mathrsfs}

\usepackage{etoolbox}
\usepackage{makeidx}
\usepackage{sectsty}
\usepackage{dsfont}
\usepackage{enumitem} 
\usepackage[]{latexsym}
\usepackage{braket}
\usepackage{caption}
\usepackage[utf8]{inputenx}
\usepackage[T1]{fontenc}
\usepackage{lmodern}
\usepackage{textcomp}
\usepackage{microtype}
\usepackage{totcount}
\usepackage{blindtext}

\newtheorem{theorem}{Theorem}

\newtheorem*{proof*}{Proof}

\newcommand{\be}{\begin{equation}}
\newcommand{\ee}{\end{equation}}
\newcommand{\bea}{\begin{eqnarray}}
\newcommand{\eea}{\end{eqnarray}}

\newcommand{\appa}{\mathscr{A}}

\newcommand{\bappa}{\mathscr{B}}

\newcommand{\trans}{\mathbf{G}}
\newcommand{\transa}{\mathbf{G}_{\mathscr{A}}}
\newcommand{\outa}{\Omega_{\mathscr{A}}}
\newcommand{\hilba}{\mathcal{H}_{\mathscr{A}}}

\title{Schwinger's Picture of Quantum Mechanics II: Algebras and Observables}

\author{F. M. Ciaglia$^{1,5}$  \href{https://orcid.org/0000-0002-8987-1181}{\includegraphics[scale=0.7]{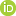}}, A. Ibort$^{2,3,6}$\href{https://orcid.org/0000-0002-0580-5858}{\includegraphics[scale=0.7]{ORCID.png}}, G. Marmo$^{4,7}$\href{https://orcid.org/0000-0003-2662-2193}{\includegraphics[scale=0.7]{ORCID.png}}\\
\footnotesize{$^{1}$\textit{ Max Planck Institute for Mathematics in the Sciences, Leipzig, Germany}} \\
\footnotesize{$^{2}$\textit{ ICMAT, Instituto de Ciencias Matem\'{a}ticas (CSIC-UAM-UC3M-UCM)}}  \\
\footnotesize{$^{3}$\textit{ Depto. de Matem\'aticas, Univ. Carlos III de Madrid, Legan\'es, Madrid, Spain}}  \\
\footnotesize{$^{4}$\textit{ Dipartimento di Fisica ``E. Pancini'', Universit\`a di Napoli Federico II, Napoli, Italy}} \\
\footnotesize{$^{5}$\textit{ e-mail: \texttt{florio.m.ciaglia[at]gmail.com}}}, \\
\footnotesize{ $^{6}$\textit{ e-mail: \texttt{albertoi[at]math.uc3m.es}}} \\ 
\footnotesize{$^{7}$\textit{ e-mail: \texttt{marmo[at]na.infn.it}}}}

\begin{document}

\maketitle

\begin{abstract}
The kinematical foundations of Schwinger's algebra of selective measurements were discussed in \cite{Ib18} and, as a consequence of this, a new picture of quantum mechanics based on groupoids was proposed.  In this paper, the dynamical aspects of the theory are analysed.  For that, the algebra generated by the observables, as well as the notion of state, are dicussed, and the structure of the transition functions, that plays an instrumental role in Schwinger's picture, is elucidated.  
A Hamiltonian picture of dynamical evolution emerges naturally, and the formalism offers a simple way to discuss the quantum-to-classical transition.
Some basic examples, the qubit and the harmonic oscillator, are examined, and the relation with the standard Dirac-Schr\"odinger and Born-Jordan-Heisenberg pictures is discussed.
\end{abstract}

\tableofcontents

\section{Introduction: Groupoids and quantum systems} 


In the previous work by the authors \cite{Ib18}, following the insight provided by J. Schwinger's in his description of Quantum Mechanical systems \cite{schwinger-the_algebra_of_microscopic_measurement, Sc70}, it was argued that the basic mathematical structure underlying the description a physical systems is that of a 2-groupoid.   

Schwinger's algebra of measurements, J. Schwinger's foundational approach to describe quantum systems and quantized fields, is based on the notion of selective and compound measurements \cite{Sc70}.   
Departing from that, Schwinger developed a theory of transitions functions that, together with a dynamical principle, set the basis to his solution of the quantum description of electrodynamics (see the celebrated series of papers \cite{Sc51}).    

After a careful analysis of Schwinger's algebra of measurements, it was argued in \cite{Ib18} that the abstract description of quantum mechanical systems should be formulated in terms of a family of primary notions: `events' or `outcomes', corresponding to elementary selective measurements; `transitions', that in Schwinger's simplified presentation were called generalised selective measurements, and `transformations', that were used to compare descriptions corresponding to different incompatible experimental setups.    

The structural properties of such notions were discussed at length, and it was shown that they have the mathematical structure known as a 2-groupoid.   
In fact, events and transitions provide a natural abstract setting for Schwinger's notion of physical selective measurements and form an ordinary groupoid.   
The theory of transformations fits naturally in this setting and determines a 2-groupoid structure on top of Schwinger's groupoid, i.e., the groupoid defined by the transitions of the system and its corresponding objects, the outcomes of the system.        

The description of the mathematical structure behind Schwinger's algebra of measurements provided in \cite{Ib18} was essentially kinematical and no attention was paid to the dynamical aspects of the theory.  
Therefore, it can be considered as a background structure for any quantum mechanical system.  
Only the broad aspects of the theory, like the role of events (but not their quantitative characteristics), the relations among them, with its categorical trait, and the inner symmetries in the form of transformations, were accounted for at this stage.

It was also shown that the fundamental representation of Schwinger's groupoid algebra allows to relate the groupoid picture to Dirac's picture of Quantum Mechanics by associating a Hilbert space to it, again reinforcing this kinematical interpretation as no dynamics in the form of a Hamiltonian operator is specified\footnote{Note that all infinite-dimensional separable Hilbert spaces are isometrically isomorphic, thus, they do not provide a distinction between quantum systems.}.  
Thus, an analysis of the fundamental dynamical aspects of the theory, starting with the notion of observable and states, should complement the work in \cite{Ib18}.  
This will be main objective of the present paper.

Here we would like to discuss in detail the role of dynamical variables, that is, physical observables, and the dynamical evolution in the groupoid setting.  
Observables will be defined in terms of the basic notion of amplitudes.  
An `amplitude' would be defined as the assignment of a complex numerical value to  any physically allowed transition of the system. 
Thus, amplitudes are just complex valued functions on Schwinger's groupoid and they will be shown to carry a $C^*$-algebra structure.  The statistical interpretation of this fundamental notion will be the subject of the forthcoming paper \cite{Ib18b}.   
The physical observables are then the real elements in this $C^{*}$-algebra.

A complete description of the system will be provided by a groupoid such that the real elements in its algebra of amplitudes are actually the totality of observables of the theory.   
In such case, the states of the theory are the states of the $C^*$-algebra of amplitudes, and their relation with vectors in the fundamental representation of the groupoid will be discussed by means of the GNS construction. 
The standard probabilistic interpretation of the theory can be established by means of the module square of amplitudes of the operators representing the observables.   

The many different, but equivalent, descriptions of the same physical system provided by (mutually incompatible) different complete families of experimental setups allow to introduce a large class of generalised transitions, called in this paper Stern-Gerlach transitions, which provide the mathematical background for Schwinger's theory of transition functions and open the path towards the formulation of a genuine dynamical principle for quantum systems.    
Some basic properties of transition functions and their dynamical properties will be analysed, however, we will leave the discussion of Schwinger's dynamical principle and its subsequent applications to be discussed elsewhere.

Before starting the actual presentation of the ideas sketched before, it is worth to devote a few lines to place the aim and scope of the present project among the many existing approaches regarding the foundations of Quantum Mechanics that could be related to it.

Apart from the standard well-known pictures of Quantum Mechanics already discussed in \cite{Ib18}, many other settings have been proposed, some of them motivated by the problem of achieving a quantum theoretical description of Gravity.   Without pretending to be exhaustive, not even covering all relevant contributions on the subject, we would like to mention here R. Penrose's spin-networks \cite{Pe70}, \cite{Ro95}, von Weizsacker \textit{urs} \cite{We58}, \cite{Ly03},  the theory of \textit{causalnets} developed from R. Sorkin's insight \cite{So97,So16}, C. Isham's categorical foundation of gravity \cite{Is03}, the noncommutative geometry approach to the description of space-time inspired on A. Connes conception of geometry \cite{Co94}, \cite{Ba10}, \cite{Do95}, etc.   All of them share a notion of ``discretness'' and ``non-commutativity'' in  Dirac's spirit \cite{Di33, Di81} towards the description of fundamental physical theories.   Even if we will not offer here a proper analysis of the relation of the present discussion with any of them, we may state that the groupoid description distilled from Schwinger's ideas is related to all of them as it describes physical systems without recurring to any \textit{a priori} notion of space-time;  moreover, this description incorporates in a natural way a statistical interpretation and may account naturally for the fundamental non-commutativity of the description of physical theories.  
  However, we must stress here that we do not pretend to use it as an alternative foundation for a `quantum' theory of gravity.

The paper will be organised as follows.  We will start by succinctly reviewing the basic notions and notations used in our previous work and, afterwards, we will discuss the properties and structure of the algebra of observables of the theory.   
The notion of a complete description of a physical system will be introduced and the $C^*$-structure of the algebra of observables will be discussed.   The notion of states and the construction of the corresponding vector descriptions in terms of the fundamental representation of the groupoid algebra will be presented by using the GNS construction.  
It will be shown that Schwinger's transition functions are naturally described in this setting, and a discussion of the properties of transition functions  will be offered.  Finally, the construction of the dynamical evolution of closed systems will be analysed proving that a Hamiltonian observable must be the infinitesimal generator of it.   Then, we will end the paper by applying all the previous ideas to discuss a few simple systems: the qubit and the harmonic oscillator.
These examples, even if elementary, illustrate the powerful analytical insight offered by  the groupoid approach. 

As it was commented before, the discussion of Schwinger's dynamical principle as well as a detailed description of the probabilistic interpretation of the theory in terms of Sorkin's quantum measures \cite{So16}, as well as the application to other physical systems of interest, will be left for subsequent works.


\section{Groupoids, algebras and other basic notions}\label{section: Groupoids, algebras and other basic notions}

Even if groupoids can be described in a very abstract setting using category theory, in this paper we will only use set-theoretical concepts to work with them.  
Thus, a groupoid $\mathbf{G}$ will be considered to be a set whose elements $\alpha$ will be called \textit{transitions} (as they represent the abstraction of actual physical transitions). 
There are two maps $s,t \colon \mathbf{G} \to \Omega$, called source and target, respectively, from the groupoid $\mathbf{G}$ into a set $\Omega$ whose elements will be called \textit{outcomes} or \textit{events} (as they are the abstraction of actual outcomes of physical measurements), and, if $s(\alpha) = a$ and $t(\alpha) = a'$, we will often use the diagrammatic representation $\alpha\colon a \to a'$ for the transition $\alpha$.  Notice that the previous notation does not imply that $\alpha$ is a map from a set $a$ into another set $a'$, even if sometimes we will use the notation $\alpha(a)$ to denote $a' = t(\alpha)$. We will  also say that the transitions $\alpha$ relates the event $a$ to the event $a'$.   

Denoting by $\mathbf{G}(a,a')$ the set of transitions relating the event $a$ with the event $a'$, there is a composition law $\circ \colon \mathbf{G}(a',a'') \times \mathbf{G}(a,a') \to \mathbf{G}(a,a'')$ such that if $\alpha \colon a \to a'$ and $\beta \colon a' \to a''$, then\footnote{The `backwards' notation for the composition law has been chosen so that the various representations and compositions used along the paper look more natural, it is also in agreement with the standard notation for the composition of functions.} $\beta \circ \alpha \colon a \to a''$.   
It is postulated that the composition law $\circ$ is associative whenever the composition of three transitions makes sense, that is: $\gamma \circ (\beta \circ \alpha) = (\gamma \circ \beta) \circ \alpha$, whenever $\alpha \colon a \to a'$, $\beta \colon a' \to a''$ and $\gamma \colon a'' \to a'''$.   

For any outcome $a\in \Omega$ there is a transition denoted by $1_a$ satisfying the properties $\alpha \circ 1_a = \alpha$, $1_{a'}\circ \alpha = \alpha$ for any $\alpha \colon a \to a'$.   
Notice that the assignment $a \mapsto 1_a$ defines a natural inclusion $i\colon \Omega \to \mathbf{G}$ of the space of events in the groupoid $\mathbf{G}$.     
Finally it will be assumed that any transition $\alpha \colon a \to a'$ has an inverse, that is, there exists $\alpha^{-1} \colon a' \to a$ such that $\alpha \circ \alpha^{-1} = 1_{a'}$, and $\alpha^{-1} \circ \alpha = 1_a$ (which expresses the fundamental physical reversibility property of transitions).  

Given an event $a \in \Omega$, we will denote by $\mathbf{G}_+(a)$ the set of transitions starting at $a$, that is, $\mathbf{G}_+(a) = \{ \alpha \colon a \to a'\} = s^{-1}(a)$.  
In the same way, we define $\mathbf{G}_-(a)$ as the set of transitions ending at $a$, that is, $\mathbf{G}_-(a) = \{ \alpha \colon a' \to a \} = t^{-1}(a)$.   
The intersection of $\mathbf{G}_+(a)$ and $\mathbf{G}_-(a)$ is the set of transitions starting and ending at $a$ and is called the isotropy group $G_a$ at $a$: $G_a = \mathbf{G}_+(a) \cap \mathbf{G}_-(a)$.     
Notice that we may write
\begin{equation}\label{G1a}
\mathbf{G} \circ 1_a = \mathbf{G}_+ (a)\, , \qquad  1_a \circ \mathbf{G} = \mathbf{G}_- (a) \, ,
\end{equation} 
in the sense that composing with the unit $1_a$ on the right selects the transitions starting at $a$.
Indeed, a transition $\alpha$ which is the result of composing some other transition with $1_a$ must have its source at $a$. 
In fact, it is easy to check that $\mathbf{G} \circ \alpha = \mathbf{G}_+(s(\alpha))$ and $\alpha  \circ  \mathbf{G}= \mathbf{G}_-(t(\alpha))$.

Given an event $a$, the orbit $\mathcal{O}_a$ of $a$ is the subset  of all events related to $a$, that is, $a' \in \mathcal{O}_a$ if there exists $\alpha \colon a \to a'$.     
Clearly the isotropy group $G_a$ acts on the right on the space of transitions with source $a$, that is, there is a natural map $\mu_a \colon \mathbf{G}_+(a) \times G_a \to \mathbf{G}_+(a)$, given by $\mu_a(\alpha, \gamma_a) = \alpha \circ \gamma_a$ (note that the transition $\gamma_a\colon a \to a$ doesn't change the source of $\alpha \colon a \to a'$).  
Then it is easy to check that there is a natural bijection between the space of orbits of $G_a$ in $\mathbf{G}_+(a)$ and the elements in the orbit $\mathcal{O}_a$, given by $\alpha \circ G_a \mapsto \alpha(a) = a'$. 
Then we may write:
$$
\mathbf{G}_+(a) / G_a \cong \mathcal{O}_a \, .
$$
It is obvious that there is also a natural left action of $G_a$ into $\mathbf{G}_-(a)$ and that $G_a\backslash \mathbf{G}_-(a) \cong \mathcal{O}_a$ too.
The subset $\mathbf{G}_+(a)$ is left-invariant under the natural action of the groupoid $\mathbf{G}$ on it, that is $\mathbf{G}\circ \mathbf{G}_+(a) = \mathbf{G}_+(a)$.  
In the same way $\mathbf{G}_-(a)$ is right invariant under the action of $\mathbf{G}$.    
Notice that if we denote by $\mathbf{G}(a)$ the union of $\mathbf{G}_+(a)$ and $\mathbf{G}_-(a)$, then  $\mathbf{G}\circ \mathbf{G}_-(a) = \mathbf{G}(a) = \mathbf{G}_+(a) \circ \mathbf{G}$, in fact, because of (\ref{G1a}), we have:
\begin{equation}\label{G1G}
\mathbf{G}\circ 1_a \circ \mathbf{G} = \mathbf{G}(a) \, .
\end{equation}

The groupoid algebra $\mathbb{C}[\mathbf{G}]$ of the groupoid $\mathbf{G}$ is defined in the standard way as the associative algebra generated by the elements of $\mathbf{G}$ with the relations provided by the composition law of the groupoid, that is, elements $\boldsymbol{\alpha}$ in   $\mathbb{C}[\mathbf{G}]$ are finite formal linear combinations $\boldsymbol{\alpha} = \sum_{\alpha \in \mathbf{G}} c_\alpha \, \alpha$, with $c_\alpha$ complex numbers. 
The groupoid algebra elements $\boldsymbol{\alpha}$ can be though as virtual transitions of the system.
Once we introduce the $C^{*}$-algebra of amplitudes in the groupoid picture, the convex combinations of the unit transitions $1_{a}$ with $a\in\Omega$ may be thought of as the normal states of the algebra of amplitudes (see Sect. \ref{sec:states} for a detailed discussion on these aspects).  
The associative  composition law on $\mathbb{C}[\mathbf{G}]$ is defined as:
$$
\boldsymbol{\alpha} \cdot \boldsymbol{\alpha}' = \sum_{\alpha, \alpha' \in \mathbf{G}} c_\alpha c_{\alpha'} \, \delta_{\alpha, \alpha'} \, \, \alpha \circ \alpha' \, , 
$$
where the indicator function $\delta_{\alpha, \alpha'}$ takes the value 1 if $\alpha$ and $\alpha'$ are composable, and zero otherwise.   
The groupoid algebra has a natural involution operator denoted $\ast$, defined as $\boldsymbol{\alpha}^\ast = \sum_\alpha \bar{c}_\alpha\,  \alpha^{-1}$, for any $\boldsymbol{\alpha} = \sum_{\alpha} c_\alpha \, \alpha$.    

If the groupoid $\mathbf{G}$ is finite, there is a natural unit element $\mathbf{1} = \sum_{a\in \Omega} 1_a$ in the algebra $\mathbb{C}[\mathbf{G}]$.   
From Eq. (\ref{G1G}) we get:
$$
\mathbb{C}[\mathbf{G}] \circ 1_a \circ \mathbb{C}[\mathbf{G}] = \mathbb{C}[\mathbf{G}(a)] \, ,
$$
with $\mathbb{C}[\mathbf{G}(a)]$ the groupoid algebra of the subgroupoid $\mathbf{G}(a)$.

Another family of relevant virtual transitions are given by $\mathbf{1}_{G_a} = \sum_{\gamma_a\in G_a} \gamma_a$, which are the characteristic `functions' of the isotropy groups $G_a$ and $\mathbf{1}_{\mathbf{G}_\pm(a)} = \sum_{\alpha\in \mathbf{G}_\pm(a)} \alpha$ that represent the characteristic `functions' of the sprays $\mathbf{G}_\pm(a)$ at $a$.  
Finally, we should mention the `incidence' or total transition, defined as $\mathbb{I} = \sum_\alpha \alpha$.   Clearly,
$$
\mathbb{C}[\mathbf{G}] \circ \mathbb{I} =  \mathbb{I}  \circ \mathbb{C}[\mathbf{G}]  = \mathbb{C}[\mathbf{G}] \, ,
$$
and
\begin{equation}
\mathbb{I} \circ 1_a =   \mathbf{1}_{\mathbf{G}_+(a)}\, , \quad  1_a \circ \mathbb{I} =   \mathbf{1}_{\mathbf{G}_-(a)}\, , \quad  1_a \circ \mathbb{I} \circ 1_a = \mathbf{1}_{G_a} \, .
\end{equation}


\section{Observables}


\subsection{The algebra of observables}

According to the premises laid on in \cite{Ib18}, we will describe a given physical system in terms of groupoids.
Specifically, we start with a family $\mathscr{A}$ of experimental setups by means of which we may perform experiments on the physical system under investigation in order to measure a physical `property' of the system.
The outcomes of such experiments are the registered `physical events' or just `events' or `outcomes' of the theoretical description of the system.   The set of all such outcomes will be denoted by $\Omega_{\mathscr{A}}$.
As in \cite{Ib18}, we will not try to make precise at this stage the meaning of `property', or the nature of the obtained outcomes as we will consider them primary notions determined solely by the experimental setting used to study our system.

The ``ontological disturbance'' of the act of measuring individuated by Schwinger is at the roots of the introduction of the notion of {\itshape transitions among the outcomes of experiments} \cite{schwinger-the_algebra_of_microscopic_measurement}:
 
\textit{``The classical theory of measurement is implicitly based upon the concept of an
interaction between the system of interest and the measuring apparatus that can
be made arbitrarily small, or at least precisely compensated, so that one can speak
meaningfully of an idealized experiment that disturbs no property of the system.
The classical representation of physical quantities by numbers is the identification
of all properties with the results of such non-disturbing measurements. It is characteristic of atomic phenomena, however, that the interaction between system and
instrument cannot be indefinitely weakened. Nor can the disturbance produced by
the interaction be compensated precisely since it is only statistically predictable.
Accordingly, a measurement on one property can produce unavoidable changes in the value previously assigned to another property, and it is without meaning to
ascribe numerical values to all the attributes of a microscopic system. 
The mathematical language that is appropriate to the atomic domain is found in the symbolic transcription of the laws of microscopic measurement}''.

In a purely classical context, the act of measuring does not influence the system and we may safely say that, if the outcome of the measurement we actually performed on the system is $a$, the measured property of the system has the value $a$. 
On the other hand, this is no longer the case for microscopic phenomena where the outcome $a$ of the  measurement of some property we actually performed on the system is compatible with different values, say, $a', a''$, etc., of the same property before the act of measurement.
The {\itshape transitions among the outcomes of experiments} (henceforth simply: {\itshape transitions}) are precisely the objects that take this instance into account.
By imposing a small set of ``natural'' axioms on it, the set $\transa$ of transitions  becomes a groupoid over the set $\outa$ of events as it is explained in \cite{Ib18}.

An \textit{amplitude} is by definition a map $f\colon \transa \to \mathbb{C}$, that is, an assignement of a complex number $f(\alpha)$ to any transition $\alpha$.   
The set $\mathcal{F}(\transa)$ of all amplitudes is an algebra with respect to the  convolution product (assuming that the groupoid $\transa$ is finite):
\begin{equation}\label{convolution}
(f\star g) (\gamma) = \sum_{\alpha \circ \beta = \gamma} f(\alpha) g(\beta) \, .
\end{equation}
where the summation is taken over all transitions $\alpha$, $\beta$ in $\mathbf{G}$ such that $\alpha \circ \beta = \gamma$.
Notice that the previous expression can also be written as:
$$
(f\star g) (\gamma) = \sum_{t(\alpha) = t(\gamma)}  f(\alpha) g(\alpha^{-1}\circ \gamma) = \sum_{s(\beta) = s(\gamma)}  f(\gamma\circ \beta^{-1}) g(\beta) \, .
$$

In general, the algebra $\mathcal{F}(\transa)$ of amplitudes is non-commutative.
However, there is a natural involution operator $\ast \colon \mathcal{F}(\transa) \to \mathcal{F}(\transa)$,  $f \mapsto f^*$, defined by:
$$
f^*(\gamma) = \overline{f(\gamma^{-1})} \, ,
$$ 
that makes $\mathcal{F}(\transa)$ into a $\ast$-algebra.

Observables are then the real elements of the algebra $\mathcal{F}(\transa)$  with respect to the involution $\ast$.
If the groupoid $\transa$ is finite, there is a unit element given by the function $\mathbf{1}$ that takes the value $1$ on all unit transitions $1_a \colon a \to a$, and zero otherwise, that is: $\mathbf{1} = \delta_{\outa}$, is the characteristic function of the set of events $\outa$ considered as a subset of $\transa$.  
Notice that:
$$
(\mathbf{1}\star f) (\gamma) = \sum_{\alpha} \mathbf{1}(\alpha^{-1}\circ \gamma) f(\alpha) = f (\gamma) \, ,
$$
and similarly $f\star \mathbf{1} = f$.
Furthermore, there is a natural norm defined on $\mathcal{F}(\transa)$ that makes it into a $C^*$-algebra\footnote{There is a natural way of constructing a $C^*$-algebra for a given groupoid over a locally compact space of events by means of a family of (left-invariant) Haar measures as described for instance in \cite{Re80} (see also \cite[Part III, Chap. 3]{landsman2007} and references therein).}.
In what follows we will assume that the algebra of amplitudes carries a $C^*$-algebra structure (see later on, Sect. \ref{sec:examples_inf}, for the explicit construction of the $C^*$-algebra of a non-finite groupoid).

In the particular instance that the groupoid $\transa$ is finite (or discrete countable), it is easy to see that $\mathcal{F}(\transa)$ is `dual'  to the groupoid algebra $\mathbb{C}[\transa]$ introduced in section \ref{section: Groupoids, algebras and other basic notions} because of the existence of  privileged bases $\{ \delta_\alpha \}$ and $\{ \alpha \} $ of the algebras $\mathcal{F}(\transa)$ and $\mathbf{C}[\transa]$ respectively, provided by the elements $\alpha$ of the groupoid itself.
Specifically, any function $f\in \mathcal{F}(\transa)$ can be written as:
$$
f = \sum_{\gamma} f(\gamma) \delta_\gamma \, ,
$$
with $\delta_\gamma$ the function that takes the value 1 at $\gamma$ and zero elsewhere.   
There is a natural pairing $\langle \cdot , \cdot \rangle \colon \mathcal{F}(\transa) \times \mathbb{C}[\transa] \to \mathbb{C}$, between the algebra of amplitudes and the groupoid algebra obtained by extending linearly the evaluation of amplitudes on transitions, that is:
$$
\langle f, \boldsymbol{\alpha} \rangle = \sum_{\alpha} f(\alpha) c_\alpha \, ,
$$
with $\boldsymbol{\alpha} = \sum_\alpha c_\alpha \alpha$.   
Under this identification the unit $\mathbf{1}$ in $\mathbb{C}[\transa]$ goes into the unit function $\mathbf{1}$ in $\mathcal{F}(\transa)$.

We may describe this identification by denoting by $\boldsymbol{\alpha}_f$ the element in $\mathbb{C}[\transa]$ associated with the function $f$ and by $f_{\boldsymbol{\alpha}}$ the function associated with $\boldsymbol{\alpha}$. 
Then, it is immediate to check that:
$$
f_{\boldsymbol{\alpha}} \star f_{\boldsymbol{\beta}} = f_{\boldsymbol{\alpha \cdot \beta}} \, , \qquad \boldsymbol{\alpha}_f \cdot \boldsymbol{\alpha}_g = \boldsymbol{\alpha}_{f\star g} \, .
$$
Moreover:
$$
\boldsymbol{\alpha}_{f^\ast} = \boldsymbol{\alpha}^\ast_f \, , \qquad f_{\boldsymbol{\alpha}^\ast} =   f^\ast_{\boldsymbol{\alpha}}  \, .
$$

It is then clear that, under suitable conditions of completeness for the norms on $\mathbb{C}[\transa]$ and $\mathcal{F}(\transa)$, the algebra of amplitudes $\mathcal{F}(\transa)$ has the structure of a von Neumann algebra because it is the dual Banach space of $\mathbb{C}[\transa]$.
This situation agrees with what happens in the algebraic formulation of quantum field theories where the relevant algebras turn out to be von Neumann algebras.


\subsection{Observables and self-adjoint operators in the fundamental representation}\label{sec:observables_fundamental}

The fundamental representation of the groupoid $\transa$ provides a natural intepretation of amplitudes in terms of operators.  
If we denote  by $\pi \colon \mathcal{F}(\transa) \to \mathrm{End}(\hilba)$ the fundamental representation of the finite groupoid $\transa$ as in \cite{Ib18}, which is given by:
\begin{equation}\label{fundamental}
\pi (f) |a \rangle = \sum_{\alpha} f(\alpha) \delta(\alpha, a) | t(\alpha) \rangle \, , 
\end{equation}
where $a \in \outa$, $|a \rangle$ denotes the corresponding vector in $\hilba$, $\delta (\alpha, a)$ is the indicator function defined as  $\delta (\alpha, a) = 1$ if $\alpha \colon a \to b$ and zero otherwise, and $t(\alpha)$ is the target of $\alpha$, i.e., $t(\alpha) = b$\footnote{There is a natural extension of this formula when the groupoid  $\transa$ is a locally compact groupoid over a standard Borel measurable space with a measure $\mu$ and a family of left-invariant Haar measures $\nu_a$.  
In such case, $\hilba = L^2(\Omega, \mu)$ and   equation (\ref{fundamental}) becomes:
\begin{equation}\label{fundamentalL2}
\pi (f) |a \rangle = \int_{s^{-1}(a)} f(\alpha) \,  | t(\alpha) \rangle \, \, d\nu (\alpha) \, .
\end{equation}}.
Note that if $\alpha \colon a \to b$, then $\pi(\delta_\alpha) |a\rangle = |b\rangle$.   Moreover. we get:
$$
\pi (f^* ) = \pi (f)^\dagger \, ,
$$
where $\pi (f)^\dagger$ is the adjoint operator of $\pi (f)$ with respect to the canonical inner product on the Hilbert space $\hilba$,
$$
\langle \psi, \phi \rangle_\Omega = \sum_{a \in \outa} \overline{\psi(a)} \phi(a)  \, .
$$
In other words, the fundamental representation is a $\ast$-representation.
Using an alternative notation $A_f = \pi (f)$, we get $A_{f^*} = A_f^\dagger$, where $A^\dagger$ denotes the adjoint operator of $A$ in the  finite-dimensional $\hilba$.   
 
Notice that $\langle b , A_f a \rangle$ is just the sum of the values of the function $f$ on the transitions $\alpha \colon a \to a'$, that is:
$$
\langle a' , A_f a \rangle = \langle a' | (A_f | a \rangle) = \sum_{\alpha \colon a \to a'} f(\alpha) \, .
$$
Notice finally that real elements in the algebra $\mathcal{F}(\transa)$, that is, functions such that $f^* = f$, are such that $A_f = A_f^\dagger$.  
In other words, real elements in the algebra of amplitudes determine self-adjoint operators on the Hilbert space $\hilba$, that is, observables in the standard framework of quantum mechanics. 
Accordingly, we call a real element in $\mathcal{F}(\transa)$ an {\itshape observable}.

For any observable $f$ we may write the following formula for the sum of amplitudes:
$$
\langle a' | A_f | a \rangle  = \sum_{\alpha \colon a \to a'} f(\alpha) \, .
$$
In the particular instance when $a = a'$, we get the real number $\langle a | A_f | a \rangle$, that can be interpreted as the expected value $\langle f \rangle_a$ of the observable $f$ in the `state' $|a\rangle$, that is:
\begin{equation}\label{expected_value}
\langle f \rangle_a = \langle a | A_f | a \rangle = \sum_{\alpha \in G_a} f(\alpha) \, .
\end{equation}

This formula justifies the name of amplitudes given before to the values of the functions $f$ on transitions.  
Actually, if there were just one transition $\alpha$ from $a$ to $a'$, like in Schwinger's measurement algebra model (see \cite{Ib18}), then the value $f(\alpha)$ is exactly the amplitude of the operator $\pi (f) = A_f$ with respect to the vectors $|a \rangle$ and $|a' \rangle$ in $\hilba$.      


\subsection{Completeness of systems of compatible observables}

Notice that the notion of observable we have introduced is consistent with the terminology introduced from the very beginning where the events $a$ were named after the outcomes of experiments performed on the system.  
In fact, given an event $a$, if we assume for simplicity that $a$ is just a real number, there is an observable in $\mathcal{F}(\transa)$ whose expected value is $a$.   
Indeed, the observable $f_a = a\, \delta_{1_a}$ is such that $\langle a | A_{f_a} | a \rangle = a$.

So far, no assumption whatsoever has been made on the structure of the whole family of amplitudes themselves $\mathcal{A}$. 
It is possible that when we use a family $\appa$ of compatible experimental setups, the algebra of amplitudes $\mathcal{F}(\transa)$ associated with the groupoid of transitions over the space of events $\outa$, yield all amplitudes of the system.    

More formally, suppose that $\mathcal{A}$ is the family of all amplitudes of the system\footnote{This is just an idealisation of a situation that would never happen, that is, we could never know for sure if the quantities we have identified as measurable for a given system are all its physical attributes that can be measured.  
For instance, think of the spin of the electron.  When Thompson identified the electron, it was just possible to measure its position, linear momentum, angular momentum, energy and charge.  
Only much later it was realised that there was another measurable physical quantity for the electron, its spin.   
We may also consider the examples provided by the many quantum charges,  isospin, barionic charge, strangeness, etc., that have been discovered later on and which are characteristic measurable quantities of elementary particles.}.  
Then, we proceed to determine experimentally as many families of events and transitions among them as possible by selecting families of compatible experimental setups $\appa$, $\mathscr{B}$, etc.   
As it was discussed in \cite{Ib18}, these families form a groupoid $\mathbf{G}$ with total space of objects $\Omega$.    

Suppose that we select a family $\appa$ of experimental setups and its corresponding subspace of events $\{a \} = \outa \subset \Omega$.  
This choice will select a subgroupoid $\transa \subset \mathbf{G}$ consisting of those transitions $\alpha \colon a \to a'$, $a,a' \in \outa$.    
Eventually, we can consider the algebra of virtual transitions of the groupoid $\transa$ and its algebra of amplitudes $\mathcal{F}(\transa)$.   
This algebra will  be contained in $\mathcal{A}$ as it was shown before.  
It could also happen that the groupoid of transitions associated with the family $\appa$ of experimental setups we have chosen is `generic' enough so that the algebra of amplitudes $\mathcal{F}(\transa)$ is essentially\footnote{In the infinite dimensional situation we will demand that the $C^*$-algebra of amplitudes generated by $\transa$ will be dense in $\mathcal{A}$ using an appropriate topology.} the whole $\mathcal{A}$. 
Then, we will say that the family of amplitudes associated with $\appa$ is a complete\footnote{Notice that this is not the standard definition of a `complete set of compatible observables'.} family of amplitudes for $\mathcal{A}$.    

As we were discussing before, that an algebra of amplitudes is complete or not could be more an academic question than a real one, in the sense that, if we find a family such that the $C^*$-algebra of amplitudes constructed from them contains all other relevant descriptions of the system, we may consider that that  algebra is just the algebra of amplitude of the system.

In what follows, we will just assume that we have a  family  $\appa$  of experimental setups such that the algebra of amplitudes of the system is given by (the closure of) the algebra $\mathcal{F}(\trans)$ functions on the groupoid $\trans$ defined by such family.  
This is not really a simplifying assumption, as the structure of the events determined by that family could be very complicated.
However, we will often use the simplifying assumption that the space of events is discrete (or even finite) to illustrate the main ideas without having to rely on heavy technical machinery from functional analysis and operator algebras.


\section{States}\label{sec:states}


Now, we are ready to discuss  properly the notion of states for physical systems described by groupoids of transitions.   
Given that the algebra of amplitudes of the system under consideration is a $C^*$-algebra $\mathcal{A}$, that will be identified with (the closure) of the algebra of functions $\mathcal{F}(\mathbf{G})$ on the groupoid $\mathbf{G}$, we define a state $\rho$ as a state on $\mathcal{F}(\mathbf{G})$ in the sense of functional analysis.
Consequently, a state $\rho$ is a normalized positive linear functional on $\mathcal{F}(\mathbf{G})$, that is, $\rho \colon \mathcal{F}(\mathbf{G}) \to \mathbb{C}$, is a linear map such that $\rho(f^*\star f) \geq 0$, for all $f$, and $\rho (\mathbf{1}) = 1$.   
Notice that we are assuming that the $C^*$-algebra $\mathcal{A}$ is unital.  

According to the previous definition, a state is an element in the dual space of $\mathcal{F}(\mathbf{G})$, however, at least in the discrete case, $\mathcal{F}(\mathbf{G})$ is the dual of the groupoid algebra $\mathbb{C}[\mathbf{G}]$ generated by transitions, and thus we may use this to construct states in the above sense.  

For instance, we may consider the linear functional $\rho_a$ defined by the unit $1_a$, that is, $\rho_a (f) = f(1_a)$, for all $f \in \mathcal{F}(\mathbf{G})$.    Clearly $\rho_a$ is a state because $\rho_a (\mathbf{1}) = \mathbf{1}(1_a) = 1$, and,
\begin{eqnarray}\label{rhof*f}
\rho_a(f^*\star f) &=& (f^*\star f )(1_a) = \sum_{\alpha\circ \beta = 1_a} f^*(\alpha) f(\beta) = \sum_{\beta \in \mathbf{G}_+(a)} f^*(\beta^{-1}) f(\beta) \\  \nonumber &=& \sum_{\beta \in \mathbf{G}_+(a)}  \overline{f(\beta)} f(\beta) = \sum_{\beta \in \mathbf{G}_+(a)}  |f(\beta ) |^2 \geq 0 \, ,
\end{eqnarray}
where the sum above should be replaced by an integral in the continuous case.  

Thus, the events or outcomes $a$ of the system can be properly identified with states $\rho_a$, recovering in this sense Schwinger identification of outcomes with physical states.    
Even more, the value $\rho_a(f) = f(1_a)$ is just the expected value of the amplitude $f$ in the state $\rho_a$, in agreement with the interpretation provided by formula (\ref{expected_value}) in the situation that there is a unique transition $1_a \colon a\to a$.  
Notice that if the system has `inner' structure, that is, if $G_a \neq \{ 1_a \}$, then the state describing the expected value of the amplitude $f$ would be the state defined as:
$$
\rho_a^{\mathrm{inner}} (f) = \frac{1}{|G_a|}\sum_{\alpha\in G_a} f(\alpha ) \, ,
$$
or, equivalently, $\rho_a^{\mathrm{inner}} = \frac{1}{|G_a|}\sum_{\alpha\in G_a} \alpha$, which is a convex combination with weights $p_\alpha = 1/|G_a|$ of all `inner' transitions $\alpha \in G_a$. 

Given a state $\rho$, we can construct the GNS Hilbert space $\mathcal{H}_\rho$ associated with it  and the corresponding representation of the $C^*$-algebra of amplitudes $\mathcal{A}$.    Let us recall that $\mathcal{H}_\rho$ is the completion of the quotient space of the algebra  $\mathcal{A} = \overline{\mathcal{F}(\mathbf{G})}$ (assuming that the descripton of the system given by the groupoid $\mathbf{G}$ is complete) with respect to the Gelfand ideal $\mathcal{J}_\rho = \{ f \mid \rho (f^*\star f) = 0 \}$.   There is a natural inner product defined on $\mathcal{F}(\mathbf{G})/\mathcal{J}_\rho$ given by $\langle f + \mathcal{J}_\rho, g + \mathcal{J}_\rho \rangle = \rho (f^* \star g)$ whose associated norm is used to construct the desired completion.   
The algebra $\mathcal{F}(\mathbf{G})$ is represented canonically on $\mathcal{H}_\rho$ as:
$ \pi_\rho (f) (g + \mathcal{J}_\rho) = f\star g + \mathcal{J}_\rho$.     

In the particular instance when we perform the GNS construction using the state $\rho_a$ defined by the outcome $a$, we get that, because of Eq. (\ref{rhof*f}), $\rho_a (f^*\star f) = 0$ iff $\sum_\beta |f(\beta)|^2 = 0$, for all $\beta \colon a \to a'$.  
Therefore, the ideal $\mathcal{J}_{\rho_a} = \{ f \mid f(\beta) = 0, \beta \colon a \to a' \}$ is just the ideal of functions vanishing on $\mathbf{G}_+(a)$, hence, we get: 
$$
\mathcal{F}(\mathbf{G})/\mathcal{J}_{\rho_a} = \mathcal{F}(\mathbf{G}_+(a)) \, .
$$
Thus, the GNS Hilbert space $\mathcal{H}_{\rho_a}$ associated with the state $\rho_a$ is given by the set of functions\footnote{Again, in the discussion to follow, beyond the case of finite groupoids, the completion of the spaces of functions with respect the appropriate topologies, should be considered.} $\psi$ on $\mathbf{G}_+(a)$ with inner product:
\begin{equation}\label{innerGa}
\langle \phi, \psi \rangle_{\rho_a} = \rho_a (\phi^*\star \psi) =(\phi^*\star \psi)(1_a) = \sum_{\alpha \in \mathbf{G}_+(a)} \overline{\phi(\alpha^{-1})} \psi (\alpha) \, ,
\end{equation}
where, with an evident abuse of notation, we use the symbols $\phi$ and $\psi$ for both the functions in $\mathcal{F}(\mathbf{G}_+(a))$ and their extension to $\mathcal{F}(\mathbf{G})$.   

Eventually, notice that the space $\mathcal{H}_{\rho_a} = \mathcal{F}(\mathbf{G}_+(a))$ supports the GNS representation of the algebra $\mathcal{F}(\mathbf{G})$, that is, $\mathcal{F}(\mathbf{G})$ acts on it by $\pi_a(f) \psi = f \star \psi $.     

On the other hand, notice that the isotropy group $G_a$ of the unit $1_a$ is contained in $\mathbf{G}_+(a)$ and it acts on $\mathbf{G}_+(a)$ by composition on the right, that is, $\gamma_a \colon \alpha \to \alpha \circ \gamma_a$, $\gamma_a\in G_a$ and $\alpha \in \mathbf{G}_+(a)$.    

Provided  the groupoid $\mathbf{G}$ is connected \footnote{If not, we will restrict ourselves to a connected component of the groupoid that, as discussed in \cite{Ib18}, will represent a sector of the theory determined by superselection rules.}, we can easily show that: 
$$
\mathbf{G}_+(a) / G_a \cong \Omega \, .
$$    
The quotient space $\mathbf{G}_+(a) / G_a$ (that is, the space of orbits of $G_a$ in $\mathbf{G}_+(a)$) is in one-to-one correspondence with the space of events $a' \in \Omega$. The map describing such correspondence is given by $[\alpha] \mapsto t(\alpha) = a'$ if $\alpha \colon a \to a'$, and $[\alpha]$ denotes the orbit passing through $\alpha$.   The map is clearly surjective because of the connectedness assumption.  To show that it is injective, notice that $t(\gamma_a \circ \alpha) = t(\alpha)$ and if we have two transitions: $\alpha, \alpha' \colon a \to a'$, then $\alpha'\circ \alpha^{-1} = \gamma_a \in G_a$ and $[\alpha] = [\alpha']$.


The GNS representation $\pi_a$ will not be irreducible in general, that is, the state $\rho_a$ is not pure in general.    We can see it by observing that there is a natural representation $\mu_a$ of the group $G_a$ on $\mathcal{H}_{\rho_a} = \mathcal{F}(\mathbf{G}_+(a))$ defined as follows:
$$
[\mu_a(\gamma_a) \psi ](\alpha) = \psi (\alpha \circ  \gamma_a) \, , \qquad \gamma_a \in G_a, \quad  \alpha \in \mathbf{G}(a) \, ,
$$
and $\psi \colon \mathbf{G}_+(a) \to \mathbb{C}$ is a function in $\mathcal{H}_{\rho_a}$.   Notice that the representation $\mu_a$ will not be irreducible in general and it will decompose as a direct sum of irreducible representations of $G_a$.   However, $\mu_a$ will always contain the trivial representation of $G_a$.  It will be given by the subspace of $G_a$-invariant functions in $\mathcal{F}(\mathbf{G}_+(a))$, that is, the subspace of functions of the form:
$$
\widetilde{\psi} (\alpha) = \frac{1}{\sqrt{|G_a|}} \sum_{\gamma_a\in G_a} \psi(\alpha\circ  \gamma_a) \, .
$$
Notice that this subspace, that can be denoted as $\widetilde{\mathcal{H}}_{\rho_a}$, is isomorphic to the Hilbert space $\hilba$, supporting the fundamental representation of the groupoid because these functions are invariant along the orbits of $G_a$, so that they project to functions on $\mathbf{G}(a)/ G_a \cong \Omega$.   The precise assignment is given by $\widetilde{\psi} \mapsto \psi$, with $\psi(a') = \widetilde{\psi}(\alpha)$, and $\alpha \colon a \to a'$.

Eventually, notice that, because of Eq. (\ref{innerGa}), we get:
$$
\langle \widetilde{\phi}, \widetilde{\psi} \rangle_{\rho_a} = \sum_{\alpha \in \mathbf{G}_+(a)} \overline{\widetilde{\phi}(\alpha^{-1})} \widetilde{\psi} (\alpha) =  \frac{1}{|G_a|} \sum_{a'\in \Omega} \sum_{\gamma_a \in G_a} \overline{\phi}(a') \psi(a') = \langle \phi, \psi \rangle_\Omega \, ,
$$
which shows that the trivial irreducible component $\widetilde{\mathcal{H}}_{\rho_a}$ of the GNS representation $\mathcal{H}_{\rho_a}$ associated with the state $\rho_a$ is isomorphic to the fundamental representation of the algebra of observables of the groupoid $\mathbf{G}$.   We can summarise the results obtained so far in the following theorem:

\begin{theorem}\label{thm: physical systems and groupoids} Given a physical system described by its groupoid $\mathbf{G}$ of transitions, and such that the unital $C^*$-algebra $\mathcal{A}$  of the system is the closure of the algebra of amplitudes $\mathcal{F}(\trans)$, there is a Hilbert space associated with the system which is provided by the Hilbert space $\mathcal{H}_\Omega$ supporting the fundamental representation of the groupoid $\mathbf{G}\rightrightarrows \Omega$.   If the groupoid is discrete countable, there is a canonical orthonormal basis $|a\rangle$, $a \in \Omega$, of $\mathcal{H}_\Omega$.

Moreover, the states $\rho_a$ determined by the unit transitions $1_a$, $a \in \Omega$ being an arbitrary outcome of the system, are naturally identified with the vectors $|a \rangle \in \mathcal{H}_\Omega$. 
The Hilbert space $\mathcal{H}_\Omega$ is isomorphic to the subspace supporting the trivial representation of the isotropy group $G_a$ in the Hilbert space $\mathcal{H}_{\rho_a}$ obtained by the GNS construction associated with the state $\rho_a$.    
\end{theorem}


\section{Schwinger's transition functions: A first approach}


The assumption that we can construct the algebra of observables of the system out of a complete family of compatible experimental setups and its corresponding groupoid of transitions, leads to some relevant observations regarding the nature and composition properties of transitions.   

Consider two complete families of experimental setups $\appa$ and $\bappa$ for a given physical system. Clearly,   $\appa$ and $\bappa$ provide two different descriptions of its family of observables $\mathcal{A}$ given, respectively, by the (closures of the) algebras $\mathcal{F}(\mathbf{G}_\appa)$ and $\mathcal{F}(\mathbf{G}_\bappa)$.  
In the case where the physical reality described by observers using the experimental setting $\appa$ cannot be different from that described by other observers using $\bappa$, we postulate that the algebras  $\mathcal{F}(\mathbf{G}_\appa)$ and $\mathcal{F}(\mathbf{G}_\bappa)$  must be isomorphic\footnote{In general, the same set of preparation procedures of a physical system may lead to the description of `different physical realities' depending on the family of experimental setups chosen to analyze it.
For instance, the silver atoms emitted by an oven, if analysed by means of a Stern-Gerlach apparatus, will lead to the description of the spin degrees of freedom of the atoms, while, if analysed by means of position detectors, will lead to the description of the localization degrees of freedom. Clearly, the task of determining whether or not two or more families of experimental setups lead to the description of `different physical realities' as above is a delicate issue that can not be solved in general, and it is thus left entirely up to the physicist analysing a concrete instance.}.   

Given their canonical $C^*$-algebraic structures, we will assume that they are isomophic as $C^*$-algebras.   
In fact, this assumption is based on physical grounds as the involution operator $\ast$ is the abstract notion of the adjoint operator in the fundamental representation, thus, the condition that the identification between both algebras is a $\star$-homomorphism is nothing else but demanding that the identification preserves the identification of observables.   

Moreover, the condition that the identification is norm preserving is just the statement that the identification of amplitudes $f(\alpha)$ with expectation values (recall Eq. (\ref{expected_value})) is preserved.   

This equivalence between the physical realities described by using different complete families of experimental setups is a sort of `relativity principle' that has deep implications on the composition properties of transitions.  
In fact, if $\appa$ and $\bappa$ represent again two complete descriptions of the system,  the algebras generated by the transitions of both systems, that is, the algebras of the corresponding groupoids $\mathbf{G}_{\appa}$ and $\mathbf{G}_{\bappa}$ to which the corresponding algebras of observables are dual (we will assume in all what follows that the groupoids defined by the systems  $\appa$, $\bappa$, are finite or discrete countable), must be isomorphic too because of the equivalence of the algebras of observables.    
We can denote by $\tau \colon \mathbb{C}[\mathbf{G}_\bappa] \to \mathbb{C}[\mathbf{G}_\appa]$ this isomorphim and by $\tau^* \colon \mathcal{F}(\mathbf{G}_\appa) \to \mathcal{F}(\mathbf{G}_\bappa)$ the corresponding isomorphism between the algebras of observables.   

We must stress that transitions $\alpha \colon a \to a'$ are observed experimentally and they occur independently of the devices we have chosen to set our experimental setting.  However, the composition law on each groupoid $\mathbf{G}_{\appa}$ depends on the events determined by $\appa$, hence, the groupoid algebra law depends on the chosen system $\appa$.    
This implies that when observing a transition $\beta\colon b \to b'$ within the `experimental frame' provided by the system $\appa$ we do not get a yes-no answer as it would be the case when observing a transition $\alpha\colon a \to a'$ with events $a,a'$ defined by $\appa$.   
However, because of the isomorphism $\tau$ between both representation it is possible to identify the transition $\beta$ with an element in the algebra $\mathbb{C}[\mathbf{G}_\appa]$, that is:
\begin{equation}\label{betaalpha}
\tau(\beta) = \sum_{\alpha \in \mathbf{G}_\appa} c(\beta, \alpha) \alpha \, ,
\end{equation} 
for some complex numbers $c(\beta, \alpha)$.   

This decomposition of transitions $\beta$ corresponding to a given `experimental frame' $\bappa$ with respect to transitions in a different, hence necessarily incompatible, experimental frame $\appa$, is instrumental in Schwinger's construction of the algebra of measurements.  

Let us recall (see \cite{Ib18} and Schwinger's original exposition \cite{Sc70}) that `transitions' are realised in  Schwinger's algebra of measurements by means of selective measurements $M_A(a,a')$, meaning by that a device that selects the system whose outcome when measuring $\appa$ is $a$ and returns the system changed in such a way that the outcome of another measure of $\appa$ would be $a'$.   
Thus, in principle, it does not make sense to compose selective measurements $M_A(a,a')$ and $M_B(b,b')$ corresponding to incompatible systems of experimental setups (unless the events $a'$ and $b$ are equivalent, as it was observed in \cite{Ib18}).    
However, at this point, in order to develop a full algebra of measurements,  Schwinger introduces the following fundamental assumption \cite[pp. 9]{Sc70}: 

\textit{``...(selective) Measurements that we have already considered involve the passage of all systems or no systems at all between the two stages, as represented by the multiplicative numbers 1 and 0.  More generally, measurements of properties $\bappa$, performed on a system in a state $a'$ that refers to properties incompatible with $\bappa$, will yield a \underline{statistical distribution}\footnote{The underlying is ours.} of possible values.  Hence only a determinate \underline{fraction} of the systems emerging from the first state will be accepted by the second stage.  We express this by the general multiplication law:
\begin{equation}\label{compound}
M(a',b') M(c',d') = \langle b' \mid c' \rangle M(a',d') \, ,
\end{equation}
where $\langle b' \mid c' \rangle$ is a number characterizing the statistical relation between the states $b'$ and $c'$.'' 
} 

Even if at first sight this interpretation of the experimental results seems to be correct, there is a fundamental issue with it. A proper probabilistic interpretation of the fraction of the systems that will emerge in the final state should be given by a positive real number, while the numbers  $\langle b' \mid c' \rangle$ appearing in the previous expansion are complex and as such are treated in Schwinger's construction of the algebra of measurements (see for instance, Eq. (1.40) in \cite[pp. 16]{Sc70}).  
Actually they must be so because they represent amplitudes of transitions.   
It is the positive real number $ |\langle b' \mid c' \rangle|^2$ the one that provides the probabilistic interpretation and the one that is actually measured in experiments.  

Thus, we conclude that Schwinger's interpretation of the composition law for compound measurements Eq. (\ref{compound}) should be properly re-interpreted.  
A proper interpretation is provided by formula (\ref{betaalpha}) above.  
To be more precise, the fundamental property that we have established is that a given physical transition $\beta$ can be described as a linear combination with complex coefficients of transitions $\alpha\colon a \to a'$ obtained from a different complete family of experimental setups $\appa$.   
Hence given two transitions $\alpha \colon a \to a' \in \mathbf{G}_{\appa}$ and $\beta \colon b \to b' \in \mathbf{G}_{\bappa}$, we can compose them once we identify $\beta$ with an element in $\mathbb{C}[\mathbf{G}_\appa]$ (or viceversa).  

We must emphasize at this point that a proper statistical interpretation of the coefficients $\langle b' \mid c' \rangle$ as well as of the notion of amplitudes and states previously discussed will be offered in \cite{Ib18b} in terms of the notion of quantum measures introduced by R. Sorkin \cite{So94}.


\section{Dynamics}   

\subsection{A first approach to dynamics on Schwinger's groupoids: Heisenberg representation}

A dynamical description of a physical system consists in prescribing the evolution of its states.  
In our current setting (see theorem \ref{thm: physical systems and groupoids}),  states are positive normalized linear functionals $\rho$ on the $C^*$-algebra $\mathcal{A}$ generated by the  observables  of the system, where the algebra $\mathcal{A}$ is identified with the $C^*$-algebra $\mathcal{F}(\mathbf{G})$ with $\mathbf{G}$  the groupoid of transitions of the system.  
The family of states will be denoted as $\mathcal{S}(\mathbf{G})$ and is a convex set in the topological dual of $\mathcal{A}$.     

However, because of the natural duality between states and observables, instead of describing the evolution of states, we may also describe the dynamical evolution of a system by means of observables.
In particular, we will consider all those dynamical evolutions that are described as a one-parameter family of positive, normalised linear maps of the $C^*$-algebra $\mathcal{F}(\mathbf{G})$.    Actually, a positive, normalised linear map $\Phi \colon \mathcal{F}(\mathbf{G}) \to \mathcal{F}(\mathbf{G})$, induces a map $\Phi^*\colon \mathcal{S}(\mathbf{G}) \to \mathcal{S}(\mathbf{G})$, as: 
$$
\Phi^*(\rho)(f) = \rho (\Phi(f)) \, , \qquad \rho \in \mathcal{S}(\mathbf{G}) \, , f \in \mathcal{F}(\mathbf{G}) \, .
$$  
This approach is the analog of Heisenberg's picture in the current setting.

A linear map $\Phi \colon \mathcal{A} \to \mathcal{B}$ is positive if it maps the positive cone of the $C^*$-algebra $\mathcal{A}$ into the positive cone of the $C^*$-algebra $\mathcal{B}$.  Then, if $\Phi$ is positive, $\Phi^*$  maps positive linear functionals into positive linear functionals.  Finally, if $\Phi$ is normalised, that is $\Phi (\mathbf{1}) = \mathbf{1}$, it maps normalised linear functionals into normalised linear functionals, $\Phi^*(\rho) (\mathbf{1}) = \rho (\Phi (\mathbf{1})) = \rho(\mathbf{1}) = 1$. Hence, if $\Phi$ is a normalised positive linear map of the $C^*$-algebra $\mathcal{F}(\mathbf{G})$, then $\Phi^*$ maps the state $\rho$ into another state $\Phi^{*}(\rho)$ of the system.   Consequently, if $\Phi_t$ is a one-parameter family of normalised positive maps, the maps $\varphi_t := \Phi_t^* \colon \mathcal{S}(\mathbf{G}) \to \mathcal{S}(\mathbf{G})$ define a dynamical evolution on the space of states.

We will not discuss here the characterisation of positive linear maps\footnote{More precisely, we would like to consider completely positive maps, but this will be discussed elsewhere where the specific adaptation of Stinespring's and Choi's theorems to the $C^*$-algebra $\mathcal{F}(\mathbf{G})$ will be analysed.} and we will leave this discussion for later analysis.   What we want to focus our attention on is the simplest situation of dynamics of closed systems.  

A closed system is a system for which its dynamical evolution is independent of external observations.
`Observations' here refers to the collection of actions undertaken by specific observers when preparing and analysing the system.   Of course, when measurements are performed, the states of the system can be modified and consequently the subsequent evolution of the states changes, however, no further modifications on the dynamical behaviour of the system are caused by the observers.
From the mathematical point of view, this means that the algebra of transitions and their transformations is not affected by the dynamics.  In turn, this means that the linear maps $\Phi_t$ describing their dynamics must preserve the composition of transitions, hence, they must preserve the convolution product in $\mathcal{F}(\mathbf{G})$:
$$
\Phi_t (f \star g) = \Phi_t (f) \star \Phi_t (g) \, .
$$
More generally, we may consider that evolution is described by a family $\Phi_{t_0,t}$ of linear transformations of the algebra $\mathcal{F}(\mathbf{G})$, where $t_0$ indicates a reference time chosen by the observer and, $t > t_0$, the time when the system is observed.  However, because the system is closed, its dynamical behaviour does not depend on the particular reference $t_0$ chosen by the observer, and we conclude that $\Phi_{t_0,t}$ depends only on the difference $s = t-t_0$, that is, $\Phi_{t_0,t} = \Phi_{t-t_0}$.  The family of maps $\Phi_t$ will be called the dynamical flow of the system.

On the other hand, if the system is closed, its dynamical evolution must be reversible, that is, the knowledge of the evolved states $\rho_t = \Phi_{t-t_0}^* \rho_{t_0}$ at time $t > t_0$ under the dynamic flow $\Phi_{t-t_0}$ allows to determine the original states $\rho_{t_0}$ by inverting the dynamics, that is, $\rho_{t_0} = (\Phi_{t-t_0}^{-1})^* \rho_t$.  Hence, the dynamical flow should consists of a family of invertible linear maps that, in addition, must satisfy: 
$$
\Phi_t \circ \Phi_s = \Phi_{t+s} \, .
$$ 
Thus, the dynamics is described by a one-parameter group of positive $\star$-preserving invertible  linear maps\footnote{In general, it is only a local one-parameter group of automorphisms as it is not guaranteed that $\Phi_t$ is defined for all $t$.}.

Moreover, it is natural to request that the dynamics  should preserve the real character of observables, that is, if $f^* = f$, then $\Phi_t (f)^* = \Phi_t(f) = \Phi_t(f^*)$.
Consequently, because we may write any element $f \in \mathcal{F}(\mathbf{G})$ as $f = f_1 + if_2$ with $f_a$, $a = 1,2$, real, $\Phi_t$ preserves the real character of observables iff $\Phi_t (f)^* = \Phi_t(f)^*$ for all $f$ and all $t$.    Therefore, we conclude that the dynamical flow $\Phi_t$ of a closed system should consists of a one-parameter group of automorphisms of the $C^*$-algebra $\mathcal{F}(\mathbf{G})$ \footnote{It is often requested that the flow satisfies a continuity property, typically being strongly continuous with respect to the topology of the $C^*$-algebra.}.

Notice that, if the $C^*$-algebra $\mathcal{F}(\mathbf{G})$ is unital and $\Phi$ is an automorphism, then necessarily $\Phi (\mathbf{1}) = \mathbf{1}$, and thus $\Phi$ is normalised.   Moreover, if $\Phi$ is an automorphims, we have $\Phi(f^\ast\star f) = \Phi(f)^\ast \star \Phi(f) \geq 0$ for any $f$, and thus $\Phi$ is positive.
Eventually, we conclude that every such family of automorphisms $\Phi_t$ defines a family of normalised positive maps.  


If we have a dynamical flow $\Phi_t$ on the $C^*$-algebra $\mathcal{F}(\mathbf{G})$, its infinitesimal generator $D$  defined as:
$$
D f = \frac{d}{dt} \Phi_t(f) \mid_{t = 0} \, ,
$$
is a derivation $D$, in principle only densely defined, i.e., $D$ it is a linear map such that $D(f\star g) = Df \star g + f \star Dg$ for all $f,g$ in the domain of $D$.   Moreover, the derivation $D$ is a $*$-derivation, that is $D(f^*) = (Df)^*$, hence, it maps real observables into real observables.   It is easy to check that given $k \in \mathcal{F}(\mathbf{G})$, the operation $D_k f = [f,k] = f\star k - k\star f$ is a derivation, moreover if $k$ is imaginary, that is $k^* = -k$, then it defines a $*$-derivation as the following computation shows:
$$
(D_k f)^* = [f,k]^* = k^*\star f^* - f^* \star k^* = f^*\star k - k\star f^* = [f^*,k] = D_k (f^*) \, .  
$$
We may assume in what follows that the derivation $D$ is bounded (what always be the case in finite dimensions) even if this will not be the case in general (see later on, Sect. \ref{sec:examples_inf}).    Moreover, if the algebra $\mathcal{F}(\mathbf{G})$ is semisimple, as it happens in the finite-dimensional case \cite{Ib18c}, the derivation $D$ will be inner, this means that there will exist an imaginary element $\tilde{h} = ih$ ($h$ real) such that:
$$
D = i[\cdot , h] \, .
$$
We will call the real observable $h$ the Hamiltonian generator of the dynamical flow $\Phi_t$ and it will determine the dynamics of the system.


\subsection{The Hamiltonian formalism}

Suppose that a Hamiltonian $h$ is given, then, we may write down the equation for the dynamics of the system in Heisenberg form as:
\begin{equation}\label{heisenberg}
\frac{d}{dt} f = i[f,h] \, ,
\end{equation}
meaning that, given an initial observable $f_0$, a solution of Eq. (\ref{heisenberg}) is a curve $f(t)$ of observables such that $df(t) /dt = i[f(t),h]$.  Because the derivation $D_h = [\cdot, h]$ is bounded, we may build its associated dynamical flow as:
$$
\Phi_t f = \exp ( {it D_h}) f = \sum_{k\geq 0} \frac{(it)^k}{k!} D_h^k(f) \, ,
$$ 
and, after some simple computations, we get that the solution to Eq. (\ref{heisenberg}) with initial value $f_0$ is given by
$$
f(t) = e^{it D_h} f_0 = \Phi_t (f_0) \, .
$$
which justifies the opening statement of this paragraph.  

We should stress that, because the fundamental representation $\pi$ is a representation of the algebra  $\mathcal{F}(\mathbf{G})$, we have $\pi(f\star g) = \pi(f) \pi (h)$, and then Eq. (\ref{heisenberg}) becomes Heisenberg's evolution equation in the standard formalism of operators in Hilbert space, that is:
\begin{equation}\label{heis_fundamental}
\frac{d}{dt} A = i [A, H] \, .
\end{equation}
where $H = A_h = \hat{h} = \pi (h)$, is the self-adjoint operator on $\mathcal{H}_\Omega$ representing the Hamiltonian $h$, and $A = A_f$ for some $f$.  Notice that any operator $A$ is the image under $\pi$ of some element $f$ in $\mathcal{F}(\mathbf{G})$.
 In particular, equation (\ref{heis_fundamental}), describes the evolution of density operators (`mixed states'), i.e., self-adjoint, non-negative, normalized operators:
\begin{equation}\label{neumann}
\frac{d}{dt} \hat{\rho} = i [\hat{\rho}, H] \, .
\end{equation}
This equation is also known as Landau-von Neumann's evolution equation.  


\section{Some simple examples and an application}


\subsection{The extended singleton and the qubit}\label{sec:examples_fin}

 We will start the discussion of examples by considering what is arguably the simplest non-trivial groupoid structure. We call it the extended singleton, and is given by the diagram in Fig. \ref{singleton} below:

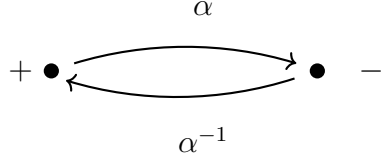
\begin{figure}[h]
\centering
\begin{tikzpicture} 
\fill (0,0) circle  (0.1);
\fill (3.5,0) circle  (0.1);
\draw [thick,->] (0.3,0.1) arc (107:73:5);
\draw [thick,->] (3.2,-0.1) arc (288:253:5);
\node [left]  at  (-0.1,0)   {$+$};
\node [right] at (3.9,0)   {$-$};
\node [above] at (2,0.6)   {$\alpha$};
\node [below] at (2,-0.6)   {$\alpha^{-1}$};
\end{tikzpicture}
\caption{The extended singleton.}
\label{singleton}
\end{figure}

This diagram will correspond to a physical system described by a complete family of experimental setups $\appa$ producing just two outputs, denoted by $+$ and $-$, and with just one transition $\alpha \colon + \to -$ among them.   Notice that the groupoid $\mathbf{G}_{\appa}$ associated with this diagram has 4 elements $\{ 1_+, 1_-, \alpha, \alpha^{-1}\}$, and the space of events is just $\outa= \{+,-\}$.  The groupoid algebra is a complex vector space of dimension 4 generated by $e_1= 1_+$, $e_2= 1_-$, $e_3 = \alpha$ and $e_4= \alpha^{-1}$, with structure constants given by the relations:
\begin{eqnarray*}
&& e_1^2 = e_1\, , \quad  e_2^2 = e_2 \, , \quad e_1 e_2 = 0 \, , \quad e_3 e_4 = e_1\, , \\ && e_4 e_3 = e_2\, , \quad e_3 e_3  = e_4 e_4 = 0\, , \quad e_1e_3 = e_3\, , \\&& e_4 e_1 = e_4\, ,  \quad e_1 e_4 = 0 \, , \quad e_3e_2 = e_3\, , \quad e_2e_3 = 0 \, .
\end{eqnarray*}

The fundamental representation of the groupoid algebra is supported in the 2-dimensional complex space $\mathcal{H} \cong \mathbb{C}^2$ with canonical basis $|+\rangle$, $|-\rangle$.  The groupoid elements are represented by operators acting on the canonical basis as:
$$
A_+ |+\rangle = \pi (1_+) |+\rangle = |+\rangle \, , \qquad  A_+  |- \rangle =  \pi(1_-) |- \rangle = 0 \, ,
$$
that is, with associated matrix:
$$
A_+ = \left[ \begin{array}{cc} 1 & 0 \\ 0 & 0\end{array}\right] \, .
$$
Similarly we get:
$$
A_- = \pi(1_-) = \left[ \begin{array}{cc} 0 & 0 \\ 0 & 1\end{array}\right] \, , \quad A_\alpha =  \pi(\alpha) = \left[ \begin{array}{cc} 0 & 0 \\ 1 & 0\end{array}\right] \, , \quad A_{\alpha^{-1}} = \pi(\alpha^{-1}) =  \left[ \begin{array}{cc} 0 & 0 \\ 1 & 0\end{array}\right] \, , 
$$
Thus, the groupoid algebra can be naturally identified with the algebra of $2\times 2$ complex matrices $M_2(\mathbb{C})$ whose  fundamental representation is provided by the matrix-vector product of matrices and 2-component column vectors of $\mathbb{C}^2\cong \mathcal{H}$.   

Amplitudes are maps $f \colon \mathbf{G}_\appa \to \mathbb{C}$, thus, they assign a complex number to any of the transitions above.   We can extend them linearly to define elements in the dual of the algebra of the groupoid $\mathbb{C}[\mathbf{G}_\appa] \cong M_2 (\mathbb{C})$.   The dual of the groupoid algebra can be identified again with the algebra of $2\times 2$ complex matrices using the standard trace inner product, that is, the inner product $\langle A , B \rangle = \mathrm{Tr\,} (A^\dagger B)$.   In particular, under this identification, observables correspond to $2\times 2$ Hermitean matrices:
\begin{equation}\label{hermitean}
A = \left[ \begin{array}{cc} x_0 + x_3 & x_1 - ix_2 \\ x_1 + ix_2 & x_0 -x_3\end{array}\right] = x_0 \, \mathbf{I} + \mathbf{x}\cdot \boldsymbol{\sigma} = \langle x , \sigma \rangle \, ,
\end{equation}
where $\sigma_\mu$, $\mu = 0,1,2,3$, denote the standard Pauli $\sigma$-matrices:
$$
\sigma_1 = \left[ \begin{array}{cc} 0 & 1 \\ 1 & 0\end{array}\right] \, , \quad \sigma_2 = \left[ \begin{array}{cc} 0 & -i \\ i & 0\end{array}\right] \, , \quad \sigma_3 = \left[ \begin{array}{cc} 1 & 0 \\ 0 & -1\end{array}\right] \, , 
$$
together with $\sigma_0 = \mathbf{I}$, and $\mathbf{x}$ is the vector in $\mathbb{R}^3$ with components $(x_1, x_2, x_3)$.
Then, the real observable $f$ defined by the Hermitean matrix $A$ above, Eq. (\ref{hermitean}), is given by:
\begin{eqnarray}
f_+ &=& f (1_+) = x_0 + x_3 \, , \quad f_- = f (1_-) = x_0 - x_3 \, , \label{f+f-}\\
 f_\alpha &=& f (\alpha) = x_1 + ix_2 \, , \quad f_{\alpha^{-1}} = \overline{f_\alpha} =  f (\alpha^{-1}) = x_1 -i x_2 \, . \label{fafa-1}
\end{eqnarray}
States  are  normalised, positive linear functionals on $M_2(\mathbb{C})$, and they can be identified with density matrices $\hat\rho = \hat\rho^\dagger$, $\mathrm{Tr\, } \hat\rho = 1$, $\hat\rho \geq 0$.   

In this representation, the complete system of observables $\appa$ will consist of the operator $\sigma_3$, identified for instance with the third component $S_z$ of the spin operator $\mathbf{S}$ of an electron.  The outcomes of this operator would be its eigenvalues $\pm1$ (that we have represented by the symbols $+$ and $-$ respectively).   Notice that in the symbolic notation used above, this observable, denoted now as $f_3$, would be defined as $f_3(1_+) = 1$, $f_3(1_-) = -1$, and zero otherwise.

Stern-Gerlach transitions will be obtained by considering another complete system of experimental setups.  After a minute reflection we will arrive to the conclusion that any other such complete system, call it $\bappa$, will also provide exactly two outcomes, we may denote them as $\{ \rightarrow, \leftarrow\} $.   
The algebra of transitions will be generated by $1_ \rightarrow$, $1_\leftarrow$, $\beta$ and $\beta^{-1}$, with $\beta$ the `flip' transition from the event $\rightarrow$ to the event $\leftarrow$.   The algebra of transitions generated by $\bappa$ will be isomorphic to the algebra of transitions generated by $\appa$, this means that there is an isomorphism $\Phi$ from the $C^*$-algebra of $2\times 2$ matrices into itself.  This isomorphism $\Phi$ will  necessarily have the form $\Phi(A) = U A U^\dagger$ with $U$ a unitary operator\footnote{A much harder problem appears when we are not considering complete descriptions, then, the map between both algebras will be just positive and we will use Choi's characterization of such transformations.}.    Notice that in such case the image of $1_\rightarrow$ in the description provided by $\appa$ will be given by $\Phi(1_b) = UA_+ U^\dagger$.

Eventually, we may consider the most general Hamiltonian dynamic for the extended singleton. For that, we may consider a general hamiltonian $H$ provided by a Hermitean matrix :
$$
H = \left[ \begin{array}{cc} h_0+h_3 & h_1 - ih_2 \\ h_1 + ih_2 & h_0 - h_3\end{array}\right] 
$$
and the evolution equation (\ref{heisenberg}) becomes, with $f_\pm$, $f_\alpha$, $f_{\alpha^{-1}} = \overline{f}_\alpha$ as in Eqs. (\ref{f+f-})-(\ref{fafa-1}):
\begin{eqnarray}
&& \dot{f}_+ = i(f_{\alpha^{-1}} h_z - \overline{h_z} f_\alpha ) \, , \\ 
&& \dot{f}_- = i(\bar{h}_z f_\alpha -  \overline{f_{\alpha}} h_z ) \, ,  \\
&& \dot{f}_\alpha = i(( f_{-} - f_{+}) h_z - 2 h_3 f_\alpha ) \, , \\ 
&& \dot{f}_{\alpha^{-1}} = i((f_+ - f_-) \overline{h_z} - 2 h_3 \overline{f_\alpha} ) \, .
\end{eqnarray}
with $h_z = h_1 + ih_2$.   Notice that $d/dt (f_+ + f_-) = 0$, i.e., the trace of $f$ is conserved.   In particular, if $f$ where a density operator $\hat{\rho}$ the trace would be preserved (and equal to 1).  

If $h_z = 0$, that is, if $H$ is diagonal, then $\dot{f}_\pm = 0$ and $f_\pm$ does not change.  If we had a classical state, that is $p = p_+ 1_+ + p_-1_-$, $p_+ + p_- = 1$, $p_\pm \geq 0$, then, for $H$ diagonal there will be no evolution of the classical state.   

Another interesting situation happens when $h_3 = 0$ and $h_z$ is imaginary, $h_z = i\nu$, $\nu > 0$.  Then, if $f_- - f_+ > 0$, we have $f_\alpha (t) \to 0$ as $t \to \infty$, thus  interpreting $f$ as measuring the amplitude of the transition $\alpha$.
In the limit of $t$ large, such amplitude vanishes. 

In the particular instance above of a classical state defined by the density operator:
$$
\hat{\rho} = \left[ \begin{array}{cc} p_1 & 0 \\ 0 & p_2 \end{array} \right] \, ,
$$
we obtain, given a hamiltonian of the form: 
$$
h_\epsilon = i \epsilon \frac{\gamma}{2} (\delta_\alpha - \delta_{\alpha^{-1}}) \, , \qquad \gamma > 0 \, ,
$$ 
that corresponds to $h_0 = h_1 = h_3 = 0$ and $h_2 = i\epsilon\gamma/2$, the dynamics:
\begin{equation}\label{eq:qubit_dyn}
\frac{d}{dt} \hat{\rho} = \epsilon \left[ \begin{array}{cc} 0 & (p_1 - p_2)\gamma/2 \\ (p_2-p_1)\gamma/2 & 0 \end{array} \right] \, .
\end{equation}

\subsection{The harmonic oscillator}\label{sec:examples_inf}

We will discuss now the paradigmatic example of the harmonic oscillator from the perspective of groupoids.  

\subsubsection{The groupoid $\mathbf{G}(A_\infty)$}\label{sec:A_infty}

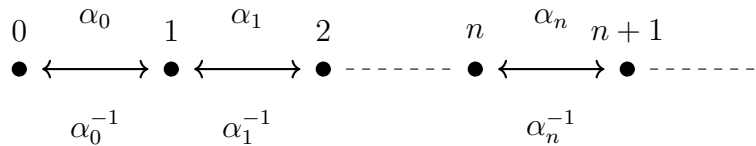
\begin{figure}[h]
\centering
\begin{tikzpicture} 
\fill (0,0) circle  (0.1);
\fill (2,0) circle  (0.1);
\draw [thick,<->] (0.3,0) -- (1.7,0);
\node at (0,0.5)  {$0$};
\node at (2,0.5)  {$1$};
\node [above] at (1,0.4)   {$\alpha_0$};
\node [below] at (1,-0.4)   {$\alpha_0^{-1}$};
\fill (4,0) circle  (0.1);
\draw [thick,<->] (2.3,0) -- (3.7,0);
\node at (4,0.5)   {$2$};
\node [above] at (3,0.4)   {$\alpha_1$};
\node [below] at (3,-0.4)   {$\alpha_1^{-1}$};
\fill (6,0) circle  (0.1);
\draw [dashed] (4.3,0) -- (5.7,0);
\node at (6,0.5)   {$n$};
\fill (8,0) circle  (0.1);
\draw [thick,<->] (6.3,0) -- (7.7,0);
\node at (8,0.5)   {$n+1$};
\node [above] at (7,0.4)   {$\alpha_n$};
\node [below] at (7,-0.4)   {$\alpha_n^{-1}$};
\draw [dashed] (8.3,0) -- (9.7,0);
\end{tikzpicture}
\caption{The diagram $K_\infty$ generating the quantum harmonic oscillator.}
\label{infinite}
\end{figure}

The kinematical description of the harmonic oscillator fits inside a family of system whose physical outcomes and transitions are described by the graph $A_\infty$, that is, the outcomes are labelled by symbols $a_n$, $n= 0,1,2,$..., and the groupoid structure is generated by the family of transitions $\alpha_n \colon a_n \to a_{n+1}$ for all natural $n$ (see Fig. \ref{infinite}).   

The assignment of physical meaning to the outcomes $a_n$ and the transitions $\alpha_n$, that is, their identification with outcomes of a certain observable and its amplitudes, will depend on the specific system under study.   

As a particular instance, we may consider that the outcomes are identified with the energy levels of a given system (the spectrum of the Hamiltonian), an atom, or the number of photons of a given frequency in a cavity for example.  

In the case of atoms, the transitions will correspond to the physical transitions observed by measuring the photons emitted or absorbed by the system.  In the case of an e.m. field in a cavity, the transitions will correspond to the change in the number of photons that could be determined by counting the photons emitted by the cavity, or by pumping a determined number of photons into it.   

At this point, no specific values have been assigned to the events $a_n$ and transitions $\alpha_n$, they just represent the kinematical background for the theory.     An assignment of numerical values to them will correspond to determining the dynamical prescription of the system.  For instance, in the case of energy levels, we will be assigning a real number $E_n$ to each event while in the case of photons, it will be a certain collection of non-negative integers $n_1,n_2,...$.     In what follows we will focus on the simplest non-trivial assignment, i.e., we assign the natural number $n$ to the event $a_n$.

The groupoid of transitions $\mathbf{G}(A_\infty)$ generated by this system is the groupoid of pairs of natural numbers or, in other words, the complete graph with countable many vertices $K_\infty$, labelled by non-negative integers, $n = 0,1,2,\ldots,$ which constitute its space of objects $\Omega = K_\infty$.    Transitions $m \to n$ will be denoted by $\alpha_{n,m}$ or just $(n,m)$ for short.  The notation in Fig. \ref{infinite} corresponds to $\alpha_n := \alpha_{n+1,n} = (n+1,n)$.

With this notation, two transitions  $(n,m)$ and $(j,k)$ are composable if and only if $m = j$, and their composition will be given by $(n,m)\circ (m,k) = (n,k)$, which corresponds to the Ritz-Rydberg combination principle of frequencies pointed out by Connes as a distinguishing ingredient of a mathematical description of quantum systems \cite{Co94}.

It is worth to point  out that the set of primary transitions $\alpha_n$ generating the graph $A_\infty$ contain all the relevant information of the system.  Any transition $\alpha_{nm}$ can be obtained composing elementary transitions: $\alpha_{nm} = \alpha_{n-1}\alpha_{n+1}\cdots \alpha_m$ ($n > m$, similarly if $n < m$).

 Notice that $(n,m)^{-1} = (m,n)$ and $1_n = (n,n)$, for all $n \in \mathbb{N}$, and that, as a set, the groupoid $\mathbf{G}(A_\infty)$ is just the Cartesian product $\mathbb{N} \times \mathbb{N}$. In what follows, we will just denote the groupoid $\mathbf{G}(A_\infty)$ as $\mathbf{A}_\infty$ for brevity.

To construct the algebra of the groupoid $\mathbf{A}_\infty$ we start by considering the set of functions which are zero except for a finite number of transitions, denoted in what follows as $\mathcal{F}_\mathrm{alg} (\mathbf{A}_\infty)$, and then we will take the closure with respect to an appropriate topology.  Thus, we may write any one of these functions as:
\begin{equation}\label{expansion_inf}
f = \sum_{n,m= 1}^\infty f(n,m) \delta_{(n,m)} \, ,
\end{equation}
where only a finite number of coefficients $f(n,m)$ are different from zero.  The function $\delta_{(n,m)}$ denotes the obvious delta function, i.e., $\delta_{(n,m)}(\alpha_{jk}) = \delta_{(n,m)} (j,k) =  \delta_{nj}\delta_{mk}$.    

The involution $f \mapsto f^*$ in the algebra $\mathcal{F}_\mathrm{alg} (\mathbf{A}_\infty)$ is defined in the standard way: $f^*(n,m) = \overline{f(m,n)}$ for all $n,m$.
Note that we may interpret functions $f$ in $\mathcal{F}_\mathrm{alg} (\mathbf{A}_\infty)$ as a formal linear combinations of elements $(n,m) \in \mathbf{A}_\infty$, that is, we can identify $\mathcal{F}_\mathrm{alg} (\mathbf{A}_\infty)$ with the (algebraic) groupoid algebra $\mathbb{C}[\mathbf{A}_\infty]$ discussed in Sect. \ref{section: Groupoids, algebras and other basic notions}.

Given two functions $f, g \in \mathcal{F}_{\mathrm{alg}} (\mathbf{A}_\infty )$ we define its convolution product $f\star g$ as the function on $\mathcal{F}_{\mathrm{alg}} (\mathbf{A}_\infty )$ whose coefficient $(n,m)$  is given by:
$$
(f \star g) (n,m)  = \sum_{(n,j) \circ (j,m) = (n,m)} f(n,j) g(j,m) =  \sum_{j} f(n,j) g(j,m) \, .
$$
Note that $\delta_{(n,m)}\star \delta_{(j,k)} = \delta_{mj} \, \delta_{(n,k)}$, where $\delta_{mj}$ is the Kronecker delta.  Moreover, $(f \star g)^\ast = g^\ast \star f^\ast$.
Hence, using Heisenberg's interpretation of observables as (infinite) matrices, we may consider the coefficients $f(n,m)$, $n, m = 0,1,\ldots,$ in the expansion (\ref{expansion_inf}) as defining an infinite matrix $F$ whose entries $F_{nm}$ are the numbers $f(n,m)$. In doing so, the convolution product on the algebra $\mathcal{F}_{\mathrm{alg}} (\mathbf{A}_\infty )$ becomes the matrix product of the matrices $F$ and $G$ corresponding to $f$ and $g$ respectively (notice that the product is well defined as there are only finitely many non zero entries on both matrices). 

The fundamental representation $\pi_0$ of the system will be supported on the Hilbert space $\mathcal{H}_\Omega$ generated by the vectors $|n\rangle$, $n = 0,1,\ldots$,  in other words, the family of vectors $\{ |n\rangle \mid  n \in \mathbb{N} \}$ defines an orthonormal basis of $\mathcal{H}_\Omega$.  Thus, the Hilbert space $\mathcal{H}_\Omega$ can be identified with the Hilbert space $l^2(\mathbb{Z})$ of infinite sequences $z = (z_0,z_1,z_2, \ldots )$ of complex numbers with $|| z ||^2 = \sum_{n = 0}^\infty |z_n|^2 < \infty$.  The fundamental representation $\pi$ is just given by (recall the definition or the fundamental representation in Sect. \ref{sec:observables_fundamental}):
$$
\pi (\alpha_{nm}) |k\rangle = \delta_{mk} |n\rangle \, ,
$$
that is, $\pi (\alpha_{nm})$ is the operator in $\mathcal{H}$ that maps the vector $|m\rangle$ into the vector $|n\rangle$ and zero otherwise or, using Dirac's notation $\pi (\alpha_{nm}) = | n \rangle \langle m |$.     

We may use the fundamental representation $\pi_0$ to define a norm on $\mathcal{F}_{\mathrm{alg}} (\mathbf{A}_\infty )$ as: $|| f || = || \pi(f) ||_{\mathcal{H}}$, and consider its completion with respect to it.   It is clear that such completion is a $C^*$-algebra because:
 $$
 || f^*\star f || = || \pi(f^* \star f) ||_{\mathcal{H}} = || \pi(f^*)\pi (f) ||_{\mathcal{H}} = || \pi(f)^\dagger\pi (f) ||_{\mathcal{H}} = || \pi(f) ||^2_{\mathcal{H}} = || f ||^2 \, .
 $$   
 Moreover, by construction, the representation $\pi$ is continuous and has a continuous extension to the completed algebra $\overline{\mathcal{F}_{\mathrm{alg}} (\mathbf{A}_\infty )}$.  By construction the map $\pi$ defines an isomorphism of $C^*$-algebras between the algebra  $\overline{\mathcal{F}_{\mathrm{alg}} (\mathbf{A}_\infty )}$ and the algebra $\mathcal{K}(\mathcal{H})$ of compact operators on the Hilbert space $\mathcal{H}$ (because compact operators are the closure in the operator norm of the subalgebra of finite rank operators).  Unfortunately, the $C^*$-algebra $\mathcal{K}(\mathcal{H})$ is too small for the purposes of describing the dynamics of a quantum system.    
 
 Then, we could proceed using the regular representation of the groupoid $\mathbf{G}(A_\infty)$ instead.
However in the case of the groupoid of pairs $\mathbf{G}(A_\infty)$ this is not strictly necessary.   We may consider the algebra of operators $\pi(\mathcal{F}_{\mathrm{alg}}(\mathbf{A}_\infty))$ as a subalgebra of the algebra $\mathcal{B}(\mathcal{H})$ of bounded operators on $\mathcal{H}$ and then, consider the von Neumann algebra generated by it, that is, its double commutant (or its closure in the weak operator topology).  It is not hard to check that it coincides with the full algebra of bounded operators on $\mathcal{H}$ because the commutant $\pi(\mathcal{F}_{\mathrm{alg}}(\mathbf{A}_\infty))'$ consists of multiples of the identity and, consequently, $\pi(\mathcal{F}_{\mathrm{alg}}(\mathbf{A}_\infty))'' = \mathcal{B}(\mathcal{H})$.  Then we conclude that the $C^*$-algebra $\mathcal{A} = C^*(\mathbf{A}_\infty)$ associated with the groupoid $\mathbf{A}_\infty$ is just the unital $C^*$-algebra of all bounded operators on the Hilbert space $\mathcal{H}$, that will be denoted in what follows by $\mathcal{A}_\infty$.

\subsubsection{The standard harmonic oscillator} 

We may define the functions $a$ and $a^\dagger$ on $\mathbf{A}_\infty$ as:
\begin{equation}\label{eq:aadagger}
a (\alpha_n^{-1}) = \sqrt{n+1} \, , \qquad a^* (\alpha_n) = \sqrt{n+1} \, , 
\end{equation}
or, alternatively, $a$ and $a^*$ are given as the formal series:
$$
a = \sum_{n = 0}^\infty \sqrt{n+1} \, \alpha^{-1}_n \, , \qquad a^* = \sum_{n = 0}^\infty \sqrt{n+1}\,  \alpha_n \, .
$$
Strictly speaking $a$, $a^*$ are not elements of the $C^*$-algebra $\mathcal{A}_\infty$, but are just functions on $\mathbf{A}_\infty$. Indeed, they define unbounded operators with a dense domain in the fundamental representation, that is, in the Hilbert space $\mathcal{H} = l^2(\mathbb{Z})$, the operators being denoted by $\mathbf{a}^\dagger = \pi(a^*)$ and $ \mathbf{a} = \pi(a)$, and they are adjoint to each other because, $\pi (a)^\dagger = \pi(a^*)$. 

Moreover, as functions on $\mathbf{A}_\infty$, we can manipulate them formally, and
a simple computation shows that:
$$
[a,a^*] = a \star a^* - a^* \star a =  \mathbf{1} \, ,
$$
with $\mathbf{1} = \sum_{n= 0}^\infty 1_n$ the unit element in $\mathcal{A}_\infty$ (note that $\pi_0(\mathbf{1}) = I$, the identity operator in $\mathcal{H}$). Then we get the standard commutation relations for the creation and annihilation operators:
$$
[ \mathbf{a} ,  \mathbf{a}^\dagger ] = I \, .
$$
Hence, we may define the Hamiltonian function:
$$
h = a^* \star a + f a^* + \bar{f} a +  \beta  =\sum_{n= 0}^\infty  n \, \delta_n  +  \sqrt{n+1} (f \alpha_n + \bar{f}\alpha_n^{-1}) +  \beta \, ,
$$
with $\omega$, $\beta$, real numbers and $f$ complex.  The corresponding equations of motion are given by:
$$
\dot{a} = i[a,h] = -i  a - i f \, , \qquad \dot{a}^* =  i[a^* ,h] = i a^* + i\bar{f}  \, ,
$$
In particular, when $f = 0$, $\beta = 1/2$, we get the Hamiltonian:
$$
h_0 = \omega \, a^*\star  a + \frac{1}{2} = \sum_{n= 0}^\infty n \delta_n + \frac{1}{2} \, ,
$$
which constitutes the standard harmonic oscillator Hamiltonian written in the abstract setting of the groupoid $\mathbf{A}_\infty$, and  with equations of motion:
$$
\dot{a} = i[a,h] = -ia \, , \qquad \dot{a}^* =  i[a^* ,h] = ia^*  \, .
$$
Using the fundamental representation $\pi$ again, the Hamiltonian operator $H_0 = \pi (h_0)$ may be identified with the Hamiltonian operator of a harmonic oscillator with creation and annihilation operators $\mathbf{a}^\dagger = \pi (a^*)$ and $ \mathbf{a} = \pi (a)$, respectively.

In addition to the creation and annihilation functions $a$, $a^*$ we may define the corresponding position and momentum functions $q$ and $p$ on $\mathbf{A}_\infty$ as:
$$
q = \frac{1}{\sqrt{2}} (a + a^*) \,, \qquad p = \frac{i}{\sqrt{2}} (a - a^*) 
$$
with commutation relations $[q,p] = i \mathbf{1}$.
Then, the canonical Hamiltonian becomes $h_0 = (p^2 + q^2) /2$.
It is interesting to observe that, by means of the fundamental representation, the groupoid functions $q,p$ become the standard position and momentum operators $\mathbf{q} = \pi(q)$, $\mathbf{p} = \pi(p)$, which are affiliated to the $C^*$-algebra $\mathcal{A}_\infty$.    It is also noticeable that other significant aspects of the harmonic oscillator, like the construction of coherent states, can also be nicely described in this setting (see for instance \cite{Co19}).


\subsection{The quantum-to-classical transition}\label{sec:classical}

As a direct application of a previous discussion, we may sketch a description of the transition from a purely quantum description of a dynamical system to a classical one.  This constitutes a relevant problem in any dynamical description of quantum systems for which there is not a general agreement on how it must be addressed.   There are many proposals and ideas on how to attack this problem (\cite{landsman2007}), some of them close in spirit to the proposal here.   A more detailed discussion of it will be pursued elsewhere.   

First of all, we shall make precise what a classical description of a physical system is.   If we have a system whose algebra of observables is given by $\mathcal{F}(\mathbf{G})$, it has a natural subalgebra provided by the functions supported on $\Omega$\footnote{Recall that $\Omega$ can be considered as a subset of $\mathbf{G}$ by using the identification of events $a$ with the units $1_a$.}, that is, the algebra of functions $\mathcal{F}(\Omega)$ that can be considered then as a subalgebra of $\mathcal{F}(\mathbf{G})$.  Notice that, if $\mathrm{supp}(f), \mathrm{supp}(g) \subset \Omega$, then, $\mathrm{supp}(f \star g) \subset \Omega$ and:
$$
f \star g = f\cdot g \, ,
$$
with $\cdot$ denoting the commutative pointwise product on functions in $\mathcal{F}(\Omega)$.  

Even more,  the representation $\pi(f)$ of a function $f$ with support in $\Omega$ is provided by the multiplication operator by the function, then $||\pi (f) || = \sup \{ || f\cdot \Psi || \mid ||\Psi || = 1\}  = \sup_{a \in \Omega} |f(a) | = || f ||_\infty$, hence, $\mathcal{F}(\Omega)$ inherits the structure of a commutative $C^*$-algebra over $\Omega$.  Thus, the commutative subalgebra of functions on $\Omega$ provides a good model for the space of observables of a classical system whose configurations are the events in $\Omega$.  On the other hand, classical states will correspond to normalized positive functional on $\mathcal{F}(\Omega)$.
For instance, if $\Omega$ is a compact topological space, then the $C^*$-algebra $\mathcal{F}(\Omega)$ becomes the $C^*$-algebra of continuous functions on $\Omega$ and the space of states the space of Radon measures on $\Omega$.   

In order to understand what kind of dynamics is induced on the classical subalgebra $\mathcal{F}(\Omega)$ from a Hamiltonian dynamics on $\mathcal{F}(\mathbf{G})$ we will assume that the Hamiltonian $h_\epsilon$ on $\mathbf{G}$ depends on a small parameter $\epsilon$ in such a way that $h_\epsilon \to h_0$ when $\epsilon \to 0$, and $h_0$ is a classical observable, that is $h_0 \in \mathcal{F}(\Omega)$.   We will be more precise on the dependence of $h_\epsilon$ in a moment.  

Notice that if $f$ is a classical observable, $f \in \mathcal{F}(\Omega)$, then, if $\alpha \colon x \to y$ is an allowed transition from $x$ to $y$, we get:
$$
[f, h_\epsilon] (\alpha) = (f(y) - f(x))h_\epsilon (\alpha)  \, ,
$$
hence, 
\begin{eqnarray*}
[f, h_\epsilon] &=& \sum_{\alpha \colon x \to y} (f(y) - f(x))h_\epsilon (\alpha) \delta_\alpha \\
&=& \sum_{\alpha \colon x \to y} f(y) h_\epsilon (\alpha) \delta_\alpha - \sum_{\alpha \colon x \to y} f(x)h_\epsilon (\alpha) \delta_\alpha \\
&=& \sum_{x \in \Omega} \left(\sum_{\alpha \in \mathbf{G}_-(x)} f(x) h_\epsilon (\alpha) \delta_\alpha - \sum_{\alpha\in \mathbf{G}_+(x)} f(x)h_\epsilon (\alpha) \delta_\alpha \right) \\
&=&  \sum_{x \in \Omega}  \left( \sum_{\alpha \in \mathbf{G}_+(x)} f(x) \bar{h}_\epsilon (\alpha^{-1}) \delta_{\alpha^{-1}} - \sum_{\alpha\in \mathbf{G}_+(x)} f(x)h_\epsilon (\alpha) \delta_\alpha \right) \\ 
&=&  \sum_{x \in \Omega} f(x) \sum_{\alpha \in \mathbf{G}_+(x)} \left(  \bar{h}_\epsilon (\alpha^{-1}) \delta_{\alpha^{-1}} -  h_\epsilon (\alpha) \delta_\alpha \right)  \, .
\end{eqnarray*}

 The quantum-to-classical transition from the quantum system $(\mathcal{F}(\mathbf{G}), h_\epsilon)$, $\epsilon > 0$, to a classical system on $\mathcal{F}(\Omega)$, will be obtained by assuming that as $\epsilon \to 0$, the amplitudes of the transitions $\alpha \colon x \to y$, $x \neq y$, tend to zero and become concentrated at the edges, $\{ x \}$ and $\{ y\}$, that is, we will assume that hamiltonian $h_\epsilon$ has a power series expansion of the form:
 $$
 h_\epsilon (\alpha ) = \epsilon h_{1,\alpha} (x,y) + \epsilon^2 h_{2,\alpha}(x,y) + \cdots \, ,  \qquad \alpha \colon x \to y \, .
 $$
 
\begin{figure}[h]
\centering
\resizebox{6cm}{5cm}{\includegraphics{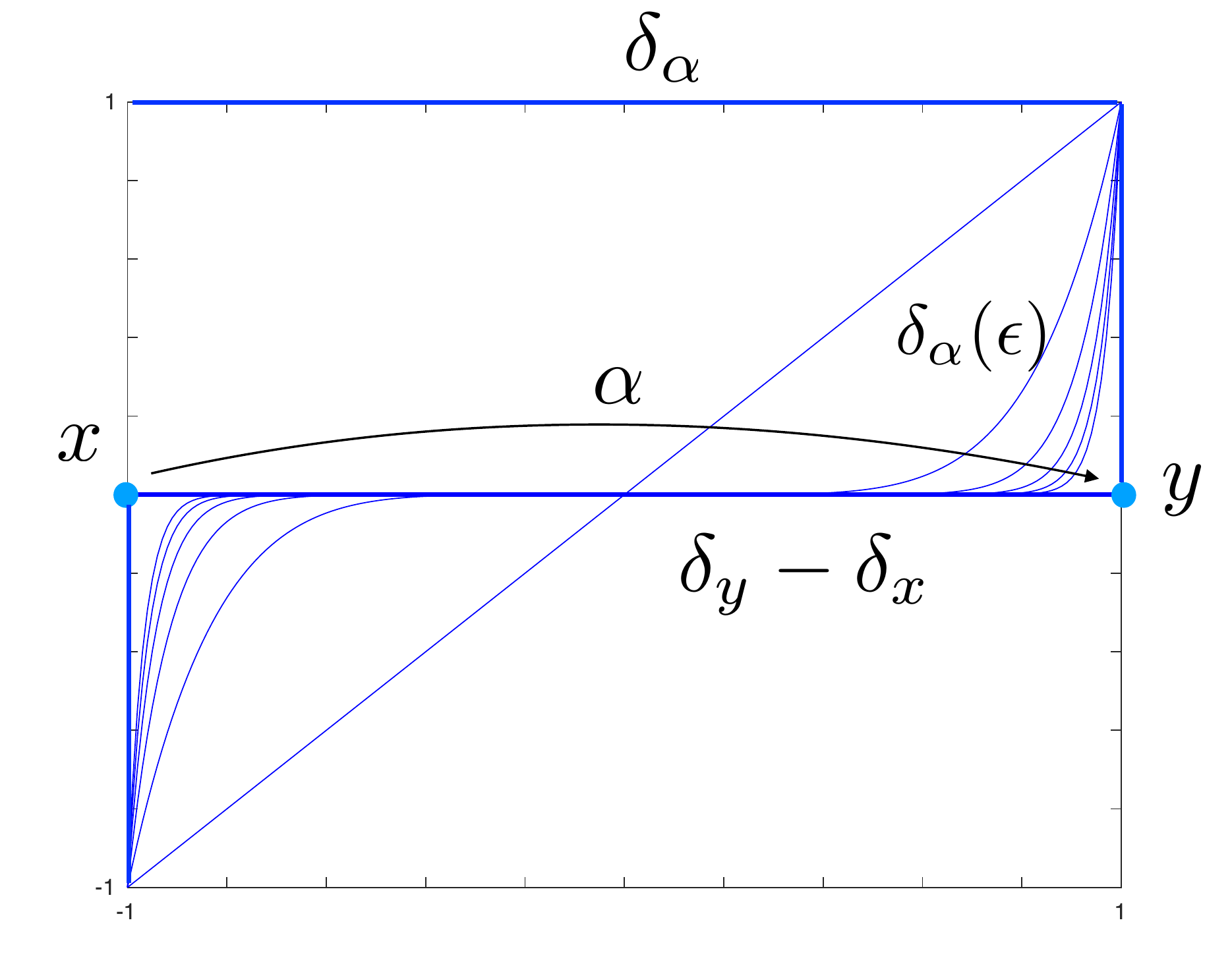}}     \caption{Quantum-to-classical deformation from $\delta_\alpha$ to $\delta_y - \delta_x$.}
\label{fig:delta_epsilon}
\end{figure}

 On the other hand, the basis functions $\delta_\alpha$ will also have to have a limit in $\mathcal{F}(\Omega)$ when $\epsilon \to 0$.  According to the previous assumption, the only natural limit for them is $\delta_y - \delta_x$ if $\alpha \colon x \to y$, or, in other words, we may imagine that there is a deformation $\delta_\alpha (\epsilon)$ such that $\delta_\alpha (1) = \delta_\alpha$ and $\delta_\alpha(0) = \delta_y - \delta_x$.   For instance, if we represent the transition $\alpha \colon x \to y$ as the oriented interval $[0,1]$, and $\delta_\alpha$ is the constant function 1, then $\delta_\alpha(\epsilon)$ could be given by the family of functions shown in the picture, Fig. \ref{fig:delta_epsilon}.
Thus we will assume that:
$$
\bar{h}_\epsilon (\alpha^{-1}) \delta_{\alpha^{-1}} -  h_\epsilon (\alpha) \delta_\alpha  = \epsilon \left( (\bar{h}_{1,\alpha} (x,y) - h_{1,\alpha}(x,y)) (\delta_y - \delta_x) \right) + h.o.t. \, ,
$$
Then, the dynamical evolution of the system is given by:
\begin{eqnarray*}
\dot{f} &=& i[f,h_\epsilon]  = \\ &=& i\epsilon \sum_{x \in \Omega} f(x) \sum_{y\in \Omega} \sum_{\alpha \in \mathbf{G}(x,y)}  \left( \bar{h}_{1,\alpha} (x,y) - h_{1,\alpha}(x,y)\right) (\delta_y - \delta_x)  \\
&= & \epsilon \sum_{x,y \in \Omega} f(x) k(x,y) (\delta_y - \delta_x)  \, ,
\end{eqnarray*}
with the kernel $k(x,y)$ given by:
$$
k(x,y) = -2 \sum_{\alpha \in \mathbf{G}(x,y)} \mathrm{Im\, } h_{1,\alpha}(x,y) \, ,
$$
and
$$
k(x,y) = - k(y,x) \, .
$$

If we consider now a change in the scale of time as $t \mapsto \tau = \epsilon t$, then the equation of motion for the classical observable $f$ becomes:
\begin{equation}\label{classical}
\frac{d}{d\tau} f = \sum_{x,y\in \Omega} ( f(x)- f(y) ) k(x,y) \delta_x \, , 
\end{equation}
or, if $\Omega$ is finite and its elements numbered $x_1, \ldots, x_n$, and the values $f(x_i) = f_i$, $k(x_i,x_k) = k_{ij}$, then:
$$
\frac{d}{d\tau} f_i = \sum_{j= 1}^n \left( k_{ij}f_j - k_{ij}f_i \right) \, .
$$
By defining the matrix $K$ with entries:
\begin{equation}\label{markov}
K_{ij} = k_{ij} - \sum_{l = 1}^n k_{il} \delta_{ij} \, ,
 \end{equation}
 we obtain:
 $$
 \frac{d}{d\tau} f = K \cdot f \, ,
 $$
 with $\cdot$ denoting the matrix vector product and $f$ denoting the column vector with entries $f_i$.  

In particular, notice that, if we consider a classical state, that is, a state of the form $p= \sum_{x\in \Omega} p_x 1_x $, $p_x \geq 0$, $\sum_x p_x = 1$, then, its evolution under a Hamiltonian function $h_\epsilon$ becomes:
\begin{equation}\label{dyn_classical}
\frac{d}{d\tau}p_i = \sum_{j= 1}^n K_{ij}\cdot p_j \, ,
\end{equation}
which has the form of the Chapman-Kolmogorov equation.
We may consider, for instance, the example from the qubit dynamics given by Eq. (\ref{eq:qubit_dyn}).  Then, by applying the quantum-to-classical deformation explained above, we obtain that the kernel $k(x,y)$ has only one entry $k_{12}$ (where the events $+,-$, are labelled now as $1,2$), therefore the $2\times 2$ Markovian matrix $K$, Eq. (\ref{markov}), becomes:
$$
K =  \left[ \begin{array}{cc} -\gamma & \gamma \\ \gamma & -\gamma \end{array} \right] \, ,
$$
and the classical dynamics of the state is given by:
$$
\dot{p}_1 = -\gamma p_1 + \gamma p_2 \, ,\qquad \dot{p}_2 = \gamma p_1 - \gamma p_2 \, .
$$

In general, because of Eq. (\ref{markov}), we get that the matrix $K$ satisfies that $\sum_{i= 1}^n K_{ij} = 0$, and then
$$
\frac{d}{d\tau} \sum_{i=1}^n p_i = 0 \, ,
$$
showing that the total probability is conserved. 
Finally, we remark that,  if $h_\epsilon$ is such that $h_\epsilon (\alpha) = i\epsilon \gamma(x,y)$ with $\gamma(x,y) < 0$, then $k(x,y) > 0$ and the classical evolution equation of the system is that of a classical random walk on the space of events $\Omega$.


\section{Conclusions and discussion}

We continued the analysis of Schwinger's formulation of Quantum Mechanics started in \cite{Ib18} by focusing on the dynamical content of the theory.
We have shown how to construct the algebra of amplitudes, and the corresponding  observables, of the theory in terms of the groupoid associated with a given physical system.
Under suitable technical conditions, this algebra is a $C^*$-algebra that we could identify with the closure of an algebra of functions on the groupoid, and we may proceed in defining the states of the theory as the normalised, positive, linear functionals on it.

In this context, every element of the space of outcomes of the groupoid of the system is realized as a normal state of the algebra, and its GNS representation is analysed.

A preliminary analysis of Schwinger's transition functions in terms of the algebra of virtual transitions is presented, leaving for further developments the analysis of their properties and their use in a dynamical context.

The description of the dynamical evolution of closed system is carried out in terms of derivations of the $C^*$-algebra of amplitudes of the system, thus, given a representation of such algebra, providing a direct connection with the Heisenberg picture of standard Quantum Mechanics.

In particular, the case of the extended singleton, and the case of the groupoid of pairs $A_{\infty}=\mathbb{N}\times\mathbb{N}$ are analysed.
Specifically, it was found that the extended singleton describes the physical system usually referred to as the qubit, while the groupoid of pairs $A_{\infty}$ may be interpreted as describing the standard quantum harmonic oscillator.

Eventually, a proposal for the description of the quantum-to-classical transition in the groupoid picture is briefly discussed.
In the context of the extended singleton, this procedure allows to obtain the classical random walk on the outcome space of the groupoid, starting from the Hamiltonian description of the dynamical evolution of the extended singleton.

Further aspects of the theory, like the role of amplitudes and the statistical interpretation of the theory, and its relation to R. Sorkin's quantum measures, or the description of non-closed dynamical evolutions in terms of Stinespring's and Choi's theorems for the von Neumann algebra associated with the groupoid of the system, will be addressed in future publications.


\section*{Acknowledgments}

The authors acknowledge financial support from the Spanish Ministry of Economy and Competitiveness, through the Severo Ochoa Programme for Centres of Excellence in RD (SEV-2015/0554).
AI would like to thank partial support provided by the MINECO research project  MTM2017-84098-P  and QUITEMAD+, S2013/ICE-2801.   GM would like to thank partial financial support provided by the Santander/UC3M Excellence  Chair Program 2018.



\end{document}